%% file: main.tex
\documentclass[preprint,journal]{vgtc}            

\onlineid{1636}

\vgtccategory{Research}

\title{How Do Researchers Manage Visualization Experiment Stimuli?}

\author{%
  \authororcid{Hyeok Kim}{0000-0003-4340-4470} and 
  \authororcid{Jeffrey Heer}{0000-0002-6175-1655} 
}

\authorfooter{
  \item
  	Hyeok Kim is with the Korea Advanced Institute of Science \& Technology, and the work was done at the University of Washington.
    Jeffrey Heer is with the University of Washington.
  	E-mail: \{hyeokk, jheer\}@uw.edu.
}

\input{sections/0-abstract}

\graphicspath{{figs/}{figures/}{pictures/}{images/}{./}} 

\usepackage{tabu}                      
\usepackage{booktabs}                  
\usepackage{lipsum}                    
\usepackage{mwe}                       
\usepackage{ccicons}                   

\usepackage{mathptmx}                  

\input{commands}

\begin{document}


\firstsection{Introduction}

\maketitle

\input{sections/1-intro}
\input{sections/2-related-work}
\input{sections/3-methods}
\input{sections/4-findings-alt}

\input{sections/5-discussion}
\input{sections/6-conclusion}

\section*{Supplementary Materials}
We provide our study protocol and qualitative codes. 

\acknowledgments{%
  The authors thank Cindy Xiong Bearfield, Danielle Szafir, and Paula Kayongo for initial feedback.
  This research was supported in part by NSF IIS Award (2402718).%
}

\bibliographystyle{abbrv-doi-hyperref}


\bibliography{reference}







\end{document}

%% file: sections/0-abstract.tex
\abstract{%
    Visualization experiments need a set of ``good’' stimuli that effectively address research questions and hypotheses.
    Creating, managing, and deploying stimuli are often challenging, as these tasks require tremendous care.
    Inappropriate stimuli can make the outcome invalid or uninteresting, wasting both researchers' and participants' resources.
    As the speed of science increases, better support for stimuli-related tasks is essential, yet we lack a closer look at how visualization researchers deal with them.
    To understand the experiences of visualization experimenters and guide future improvements, we interviewed 19 visualization researchers with diverse backgrounds and experiences.
    Our findings describe practices and challenges across the life cycle of stimuli, from exploration and selection through shipment, deployment, and analysis.
    For example, stimuli management and deployment require tedious manual effort, which does not scale for experiments with many levels and complex conditioning.
    We also discuss both concerns and optimism around AI-assisted visualization experiment design.
    We conclude with future research opportunities in supporting stimuli creation, automated stimuli inspection, and experimental apparatus concerns.
}

\keywords{Data visualization, experiment design, stimuli generation, interview}

\teaser{
  \centering
  \includegraphics[width=\linewidth, alt={There are two sections, labeled A and B, summarizing our findings. Section A describes the life cycle of stimuli through six stages: seed idea, exploration, selection, shipment, deployment, and analysis. After a formal study, the analysis outcome can be reported or fed back into the seed idea stage for a subsequent study. After pilot studies, analysis outcomes can be used to update parameters, check manipulations, and assess feasibility across the exploration, selection, and shipment stages. Section B summarizes the challenges of managing stimuli in terms of objectives and constraints. The primary objectives are thoroughness, which ensures the quality of stimuli in breadth and depth, and controllability, which ensures full and fine control over stimuli. Constraints prevent researchers from fulfilling these objectives and include labor, skill, uncertainty, and unfixable constraints. Labor constraints for the thoroughness objective include inspecting shipment on a study platform, while those for controllability include overseeing the entire stimuli and labeling real-world charts for control. Skill constraints for thoroughness include debugging for non-desktop devices and browsers and preventing or addressing technical failures, while those for controllability include creation skills for uncommon visualization designs and maintaining consistency of charts from varying sources and tools. Uncertainty constraints for thoroughness include navigating a large space of possible design options and no guidance beyond convention, while those for controllability include ensuring the demonstration of intended effects in stimuli and disagreement among collaborators in labeling. Lastly, unfixable constraints for thoroughness include researchers' time and finances as well as participants' cognitive capacity, while those for controllability include lacking effective control over synthetic data and the limited customizability of study platforms.}]{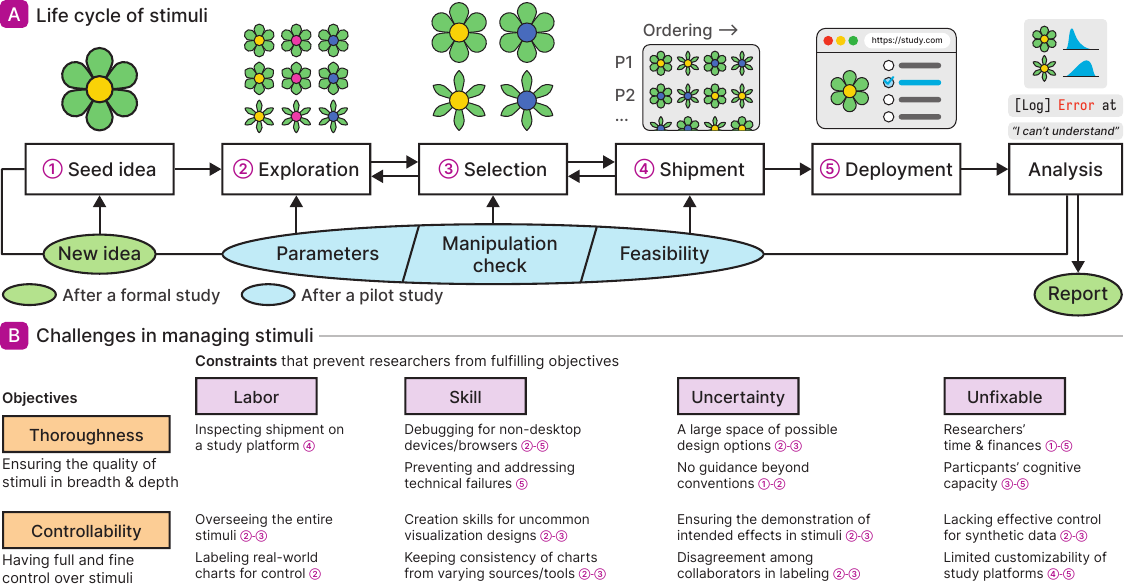}
  \caption{%
    (A) Lifecycle of visualization experiment stimuli. Rectangles indicate current status, while circles indicate decisions made after running a pilot (lightblue) or formal (green) study. The locations of decisions can vary, but here indicate where our interviewees most commonly made those decisions. 
    (B) We characterize researchers' challenges in creating and managing visualization experiment stimuli as \textit{constraints} that prevent pursuing \textit{objectives}. The listed challenges are examples.
  }
  \label{fig:teaser}
}


%% file: commands.tex
\usepackage{xspace}
\usepackage{listings}
\usepackage{array}
\usepackage{multirow}
\usepackage{soul}
\usepackage{tcolorbox}

\newcommand{\ie}{{i.e.,}\xspace}
\newcommand{\eg}{{e.g.,}\xspace}
\newcommand{\cf}{{c.f.}\xspace}
\newcommand{\ea}{{et~al.}\xspace}

\newcommand{\etc}{{etc.}\xspace}

\newcommand{\bpstart}[1]{\vspace{1mm} \noindent{\textbf{#1.}}}

\newcommand{\bstartnc}[1]{\vspace{1mm} \noindent{\textbf{#1}}}

%% file: sections/1-intro.tex
Experiment studies help us to understand how people use visualizations to perceive data, update prior beliefs, and make decisions.
Such findings provide guidelines on visualization design and also facilitate systems for design reasoning~\cite{kim2017:graphscape} and recommendation~\cite{moritz2018:formalizing,li2022:kg4vis,hu19:vizml}.
Experimenters usually design studies based on hypotheses, which are embodied as \textit{stimuli} to elicit subjects' reactions.
For example, to test the effectiveness of quantitative color scales~\cite{liu2018:colormap} or build a related theoretical framework~\cite{bujack2018:colormap}, the experimenters need to produce a set of relevant color scales.
Given its essential role of bridging research questions to human perception, experimenters need to generate ``good'' stimuli to satisfy criteria like validity and generalizability.
In the above example, the experimenter would want to test a ``decently comprehensive'' set of realistic color scales, including those commonly used.
In addition, they may need to consider other relevant factors, such as data cardinality and distribution, that could influence the results. 

However, decisions for stimuli generation are often nuanced and learned via apprenticeship, making it challenging for researchers without prior experience.
For example, some studies use more than a few dozen charts, each of which needs careful inspection, potentially posing challenges in design iteration and version control.
These challenges motivate better support for stimuli creation and management beyond those for web browser-based experiment development and deployment~\cite{nobre2021:revisit,cutler2026:revisit,deleeuw2023:jspsych}.
A first step is to closely understand researchers' challenges and strategies.
Although prior work has documented practices in crowd-sourcing experiments~\cite{borgo2018:crowd}, the curse of a large volume of stimuli~\cite{lam2012:scenarios}, and sources of miscommunication in study stimuli~\cite{nobre2024:literacy}, little research has been done on the process of designing and deploying visualization experiment stimuli. 
Next, the research community anticipates and experiences substantial influences of AI in experiment design~\cite{musslick2025:aiScience,bianchini2022:aiScience} and visualization practices~\cite{wu2022:ai4vis}.
To better inform the design of appropriate AI and other automation methods for visualization stimuli, it is timely to contextualize them within researchers' practices.
Therefore, we ask the following research questions.

\begin{itemize}
    \item RQ1. What challenges and strategies do visualization experimenters have in designing, creating, and managing stimuli?

    \item RQ2. What are promising research opportunities to support how visualization experimenters work with stimuli?
\end{itemize}

To answer these questions, we interviewed 19 visualization experimenters from 14 academic organizations with varying degrees of experiences (from Ph.D. students to senior faculty).
Our primary scope is online, crowd-sourcing experiments given its wide use in the field~\cite{borgo2018:crowd}.
We report researchers' challenges in exploring, selecting, and shipping stimuli to study platforms by characterizing them as constraints in achieving \textit{thoroughness} and \textit{controllability} objectives, as summarized in \Cref{fig:teaser}.
Thoroughness objectives refer to researchers' goals to ensure the quality of stimuli in terms of both breadth and depth. 
For example, navigating a large space of possible designs and inspecting many stimuli are example challenges in pursuing thoroughness.
Controllability objectives pertain to having full and fine control over stimuli. 
Limited customizability in stimuli design and an inability to oversee the entire stimuli set were common barriers in achieving controllability.
We then present their strategies, such as the importance of ``seeing'' stimuli ideas and thinking through statistical analysis methods as guidelines for stimuli design.
Regarding AI-assisted visualization experiments, interviewees had concerns about bias and maintaining authority, as well as optimism in automated creation and inspection.
Based on our findings, we discuss future research opportunities in developing support for stimuli creation and management, response simulation for inspection, and cross-apparatus design.

%% file: sections/2-related-work.tex
\section{Related Work}\label{sec:rw}

We refer to \textit{visualization experiments} as controlled human-subject studies where data visualizations are primary stimuli. 
Based on the categorization by Lam~\ea~\cite{lam2012:scenarios}, our scope includes studies that evaluate \textit{design}s (visual encodings and human perception \& cognition) and \textit{redesign}s (improvement for previously accepted designs).
However, our scope does not cover experiments to evaluate \textit{prototype systems} (\ie~newly proposed visual analytic or authoring tools), as these studies involve a different (although not mutually exclusive) set of challenges. 

\subsection{Visualization Experiment Design}
Through a survey of empirical visualization studies, Lam~\ea~\cite{lam2012:scenarios} characterize high-level objectives.
First, experimenters want to understand design features that affect human visual data perception and cognition (\textit{user performance}) and contribute to effective charts (\textit{user experience}).
Next, they want to test how different chart designs support different communication tasks, such as fact checking (\eg~\cite{linsnic2024:fact}) and risk communication (\eg~\cite{zhu2024:risk}) (\textit{communication through visualization}). 

Carpendale~\cite{carpendale2008:eval} contextualizes literature on social science research methods to inform challenges in quantitative visualization studies.
Among them, \emph{internal}, \emph{construct}, and \emph{ecological} validity concerns are relevant to stimuli creation.
An experiment achieves internal validity once stimuli test an effect with no confounders.
If we broadly consider stimuli, including task instructions and solicitation questions, an experiment satisfies construct validity when the tasks and stimuli address the intended effect.
An experiment holds ecological validity when stimuli can be translated to real world scenarios.
Internal and ecological validity concerns inherently form a tradeoff, where the pursuit of ecological validity can sacrifice internal validity and vice versa.
For example, using real-world visualizations as stimuli can cause confounders like color semantics, whereas participants from a general population are unlikely to see laboratory-generated charts in their everyday situations.
Our findings illustrate challenges and strategies in navigating with this tradeoff.

Crowd-sourcing platforms, such as MTurk, Prolific, and Qualtrics, allow experimenters to run studies for a larger population beyond on-campus participants in a timely manner. 
Crowd-sourcing became a norm in contemporary visualization experiments.
By analyzing 180 crowd-sourcing experiments from 82 papers, Borgo~\ea~\cite{borgo2018:crowd} point out poor reporting quality and synthesize a checklist and template for reporting, which nevertheless lack considerations regarding stimuli. 
In addition, they pose a potential bias in favoring experiment types that are feasible for platforms or do not involve technical complexity in remote settings (\eg~static visualizations).

Elliot~\ea~\cite{elliot2021:methods} offer a design space for vision science methods for visualization experiments, in terms of participant tasks, stimuli adjustment, response solicitation techniques, and performance measurements. 
Among them, stimuli adjustment is most relevant to our scope.
A given independent variable (\eg~the difference between two bar heights) can be adjusted gradually (increasing or decreasing unit by unit), interactively (participants indicate when they can tell a meaningful difference), randomly, and based on prior responses (making it more subtle after a correct response).
Through a crowd-sourced study, Nobre~\ea~\cite{nobre2024:literacy} identify literacy barriers for data visualization: human participants may misunderstand task instructions or objectives, misinterpret chart designs (\eg~visual encodings and references), or read wrong values (\eg~maximum instead of average; reading \textit{y} instead of \textit{x}). 
To mitigate such barriers, experimenters may need to choose simpler visual designs and tasks, despite some criticism against simplicity~\cite{plaisant2004:challenge}.
Alternatively, Burns~\ea~\cite{burns2020:level} propose a framework for accounting for different levels of understanding in a crowd-sourcing setting. 

While the literature documents various concerns in visualization experiment design, these are limited in their ability to provide insights about the nuanced decisions that experimenters make while creating stimuli. 
Ziemkiewicz~\cite{ziemkiewicz2020:open} only point out related gaps in our understanding, such as the number of cases to test and the lack of characterization of common tradeoffs.
Our work aims to understand what factors form challenges in creating stimuli and how researchers deal with them in their lived experience as empirical visualization researchers.

\subsection{Technical Support for Visualization Experiments}

As surveyed by Cutler~\ea~\cite{cutler2026:revisit}, researchers build and deploy crowd-sourcing visualization experiments using survey platforms (\eg~Google Forms, Qualtrics), psychology experiment toolkits (\eg~psytoolkit~\cite{stoet2017:psytoolkit}, PsychoPy~\cite{peirce2007:psychopy,peirce2019:psychopy2}, jsPsych~\cite{deleeuw2023:jspsych}), and visualization-specific toolkits (\eg~VisUnit~\cite{jianu2025:visunit}, GraphUnit~\cite{okoe2015:grapunit}).
Recently, ReVISit~\cite{nobre2021:revisit,cutler2026:revisit} helps researchers build experiment websites with customizable components, support for external libraries (\eg~Vega-Lite, custom JavaScript), diverse response methods (\eg~survey forms, audio and screen recording, drawing), and domain-specific language-based creation. 
While ReVISit~\cite{cutler2026:revisit} supports browsing individual responses, other research has focused on more general data analysis toolkits.
To highlight some recent examples, Jun~\ea~\cite{tea2019:tea,jun2022:tisane,jun2024:rtisane} offer a range of systems to automate coding statistical analyses based on conceptual hypotheses, and Lam~\ea~\cite{lam2024:lloom} propose a Large Language Model (LLM)-based tool for qualitative analysis. 
Currently, we have little understanding regarding whether we need support for stimuli creation, and if so, what kind of support the research community wants.
To better inform future tool development, we aim to uncover what forms of technical support researchers desire regarding stimuli, as well as how their practices and concerns translate to future tool development.

\subsection{AI in Experiments}
Adoption of AI for scientific research is an active topic across diverse domains. 
Given the breadth of this discourse, we only point to topics relevant to our scope: literature review, hypothesis generation, stimuli creation, experiment design, and response simulation.
Readers can refer to review papers~\cite{musslick2025:aiScience,bianchini2022:aiScience} for further reading.
Wu~\ea~\cite{wu2022:ai4vis} also provide a detailed review of the use of AI in data visualization.

First, AI shows promise to support reviewing a wide range of prior literature, with emphasis on querying related papers~\cite{arpit2022:vitality,agarwal2025:litllms} and generating and/or assisting writing of reviews~\cite{an2024:vitality2,choe2024:papers,agarwal2025:litllms}.
Second, generating hypotheses requires a well-structured, domain-specific corpus, such as molecular structures for material discovery~\cite{merchang2023:materials}.
In data visualization, Zeng~\ea~\cite{zeng2023:dataset,zeng2024:tooManyCooks} present a corpus for graphical perception studies in a principled methods. 
Kim \& Heer~\cite{kim2026:kbAug} propose a method for finding design features (\eg~the use of log scale on color) that are not covered by such corpora, which can potentially be turned into hypotheses.
Next, hypotheses about data and visualization need to be reified as stimuli. 
Prior work on automatic visualization creation tends to suppose design recommendation scenarios (\eg~\cite{yang2023:draco2,moritz2018:formalizing,wongsuphasawat2016:voyager,wongsuphasawat2017:voyager,kim2023:dupo,hu19:vizml,zhang2024:adavis,pandey2023:genorec,pandey2023:medley,wu2022:multivision}) and data-driven Q\&A settings (\eg~\cite{sah2024:nl4dvllm,kafle2018:dvqa}).

Fourth, independent, moderation, and control variables in hypotheses need to be assigned as study conditions (\eg~within, between, and randomization).
While ReVISit~\cite{cutler2026:revisit} supports enumerating condition assignments with different randomization options (\ie~the order of stimuli per participant), deciding how to assign variables remains the researchers' job given that it involves thorough, iterative reasoning about validity and study feasibility.
Lastly, simulated responses (using AI in addition to, or even instead of, human participants) have been proposed to reduce resource constraints in human-subject studies, with some successful adoptions in some settings like marketing~\cite{brand2023:marketing}.

Researchers have both hopes and concerns in using AI for scientific knowledge discovery.
Through a survey of 1,600 scientists, Van Noorden and Perkel~\cite{vanNoorden2023:survey} find that scientists are optimistic that AI can help with lowering language barriers, reducing time and financial costs, and making coding and presentation easier.
However, they also point out shared concerns, such as false information and bias in overall writing and literature review, not being able to understand underlying mechanisms, and widespread plagiarism and fraud (\eg via data fabrication).
Indeed, Hao~\ea~\cite{hao2026:focus} find that AI-assisted research papers exhibit higher publication and citation rates, but constricted topical coverage.
In social science contexts, Grossman~\ea~\cite{grossman2023:transform} identifies issues with the use of LLM for response simulation, such as biases due to reusing the same underlying distribution and reproducibility concerns of commercial models. 
Taking a more formal lens, Hullman~\ea~\cite{hullman2026:iid} note that AI response simulation violates the assumption of independence and identical distribution (IID), may fail to capture detailed distributional characteristics (\eg~variance, skewness), and provides a ``modest gain in precision'' of findings. 
In visualization, computer vision-based methods~\cite{haehn2019:cnn}, human-AI alignment~\cite{wang2025:aligned,kafle2018:dvqa}, and/or LLM-based design judgment~\cite{wang2025:dracogpt} could be used for simulated user responses.
Based on our findings, we discuss how these concerns are contextualized in visualization experiment settings. 

%% file: sections/3-methods.tex
\section{Methods}\label{sec:methods}

To understand challenges, strategies, and desired tools in dealing with visualization experiment stimuli, we interviewed 19 visualization researchers with experience in running experimental studies. 
Our study was approved by the IRB of the University of Washington (No. STUDY00024973).

\subsection{Interviewees}

We recruited 19 visualization researchers (\Cref{tab:participants}) through professional networks (research Slack groups, LinkedIn, and Bluesky) and by directly contacting those who recently published experiment research.
To solicit diverse ideas, we carefully recruited them by diversifying their academic organizations and their experience levels.
Interviewees were from 14 different organizations, and the maximum number of interviewees from the same institute was three.
Our interviewees' experience levels were reasonably distributed: nine junior (Ph.D. students), five early career (postdocs, assistant professors), and five senior (tenured professors) researchers.  
Their research areas collectively covered perception \& cognition, decision making, communication, data/visualization literacy, equity \& accessibility.
Our recruitment criteria were: (1) having completed visualization experiment studies and (2) being able to communicate in English (our study language).

\input{figs/tab-participants}

\subsection{Procedures}

We conducted semi-structured interviews. 
Each session took about 45 minutes and was conducted in English through a remote meeting platform (Teams).
We prepared our interview questions after three pilot sessions (one junior, one early career, and one senior researcher). 
After obtaining verbal consent, we asked questions about their years of experience running visualization experiments and elicited keywords for their research background.
To gain context on their research practices, we asked about their most recent experiment study experience. 
For faculty interviewees, we asked them to describe the study they could best remember.
Then, we inquired about their challenges in designing experiment stimuli, the impact of stimuli design on data analysis, and how they managed such challenges.
We asked follow-up questions regarding aspects that either commonly or rarely appeared in prior sessions to reason about consistency and diversity across their experiences.
Finally, we asked them to brainstorm technological support for experiment stimuli design, as well as tasks that they did not want to replace with automation technologies. 
Interviewees were compensated with a USD \$20 gift card. 
Supplementary Material includes the detailed interview protocol.

\subsection{Analysis}

We collected audio recordings from the sessions. 
We analyzed the recording transcripts through thematic analysis~\cite{braun2006:thematic}.
Initially we performed inductive open coding of the interview transcripts, identifying codes like `language-driven confounders,' `using custom tools for inspection,' and `novelty in task design.'
Then, guided by our research questions, we deductively categorized the open codes into themes regarding challenges, strategies, and desired tools in stimuli creation and management in visualization experiments.
This step resulted in initial themes, such as `making decisions for using real-world or synthetic stimuli' and `expanding and pruning stimuli.'
To build a more cohesive structure between those themes and better illuminate future research opportunities, we reorganized our themes in terms of objectives and constraints in dealing with stimuli.
We include the evolution of our coding process in Supplementary Material.

%% file: figs/tab-participants.tex
\begin{table}[tb]
  \caption{%
  	The demographics of interviewees. Exp. Levels (Experience levels): Junior (Ph.D. students), Early career (postdocs, assistant professors), and Senior (tenured professors). Years. (Years of experience in visualization experiments, not entire research career). P\&C (perception \& cognition), DM (decision making), E\&A (equity \& accessibility), Comm. (communication). The creation modes of our interviewees reflect only the experiences mentioned in their sessions. %
  }
  \label{tab:participants}
  \scriptsize%
	\newlength{\digitwidth}%
  \settowidth{\digitwidth}{0}%
  \centering%
  \begin{tabu}{%
  	  l%
  	  	l%
        l%
        l%
        l%
  	}
  	\toprule
    \textbf{\small{Participant}} & \textbf{\small{Exp. Levels}} & \textbf{\small{Years}}  & \textbf{\small{Broad areas}} & \textbf{\small{Creation modes}} \\
  	\midrule
P1	& Junior & 3-5 & P\&C & Synthetic \\
P2	& Junior & 0-2 & DM, Comm. & Synthetic\\
P3	& Junior & 3-5 & P\&C  & Synthetic \\
P4	& Early Career & 5-10 & E\&A, Comm. & Synthetic \\
P5	& Early Career & 5-10 & P\&C & Real-world \\
P6	& Early Career & 10-15 & P\&C & Both \\
P7	& Junior & 3-5 & P\&C, DM & Synthetic \\
P8	& Junior & 3-5 & P\&C & Synthetic \\
P9	& Senior & 16-20 & P\&C & Both \\
P10	& Senior & 10-15 & DM, literacy & Both \\
P11	& Senior & 16-20 & P\&C  & Synthetic \\
P12	& Early Career & 10-15 & P\&C, Comm. & Both \\
P13	& Junior & 3-5 & P\&C, Comm. & Both \\
P14	& Early Career & 3-5 & DM, Comm. & Real-world \\
P15	& Junior & 3-5 & Literacy & Synthetic \\
P16	& Junior & 0-2 & DM & Real-world \\
P17	& Junior & 0-2 & P\&C, E\&A & Real-world \\
P18	& Senior & 16-20 & P\&C & Synthetic \\
P19 & Senior & 10-15 & Literacy, design & Real-world \\
  	\bottomrule
  \end{tabu}%
\end{table}


%% file: sections/4-findings-alt.tex
\section{Findings}
We first describe challenges and strategies occurring while creating and managing visualization experiment stimuli.
We then report concerns and opportunities in adopting AI for stimuli.
Below, ``interviewee'' strictly refers to those who participated in our study; ``participant'' and ``user'' refer to those who took part in our interviewees' experiments.

\bpstart{Privacy statement}
Describing detailed information about our interviewees' experiments---which often cover niche topics studied by a limited number of people---can inadvertently reveal their identities especially when this information is combined with interviewees' demographics. 
To protect the privacy of our interviewees, we only provide background information that is minimally necessary for understanding their challenges and answering our research questions. 
We also do not identify whether two or more interviewees were from the same institute.

\subsection{Life Cycle of Stimuli}

To contextualize our findings, we start with an overview of decision making steps that our interviewees made while dealing with stimuli, as summarized in \Cref{fig:teaser}A.
First, there is a \textbf{seed idea} about the stimuli, which could be from real-world experiences, examples, or demands (P2, P4, P9, P10, P16, P18), a specific prior paper on similar problems (P3, P6, P9, P14, P15), or what is absent in a broader theoretical framework (P8).
Second, experimenters go through an \textbf{exploration} phase to populate more examples. Initial examples (typically one or two) may not be sufficient to test intended effects with a reasonable degree of generalizability.
In this phase, interviewees add or remove independent variables (P2, P11) or increase or decrease levels of each variable (P1, P15).
When using real-world examples, they often felt the need to include diverse examples or data patterns (P5, P10, P13, P17).
Third, they make a \textbf{selection} of final stimuli to use in experiments.
This selection step can provide a reasonable set of constraints and hence facilitate further exploration.
For example, P11 initially considered using various chart types, which exploded the number of stimuli with no relationship between them. 
Once fixing it to bar charts, they were able to explore possible variations within that scope with no explosion.

Next, the selected stimuli are \textbf{shipped} to a crowd-sourcing platforms like Prolific or Qualtrics (P8, P12, P17), a study website using ReVISit~\cite{nobre2021:revisit,cutler2026:revisit} (P3, P9, P14, P16), or custom websites (P9, P10, P16).
While our interviewees generally run experiments via crowd-sourcing, a few of them also mentioned prior or on-going experiences in running in-person studies (P9, P18) using websites or instruments.
Here, they need to check if stimuli are correctly assigned to corresponding conditions, if the randomization is working as intended, and if there are any technical bugs, for example.
Once ready, experimenters \textbf{deploy} their experiment and wait for responses.
To avoid confusion with visualized data within stimuli, we use \textit{response} to refer to collected user reactions.
Finally, experimenters \textbf{analyze} the collected responses. 
In this step, stimuli are revisited to match the findings.
Note that these steps are not necessarily linear, as prior steps may be revisited back and forth.

This cycle applies not only to final formal studies, but also pilot studies. 
Our interviewees ran multiple pilot studies for various purposes, including parameterization of stimuli (\eg~correlation, difficulty), manipulation checks (\eg~natural language consistency, the existence of effects), feasibility (\eg~time and load on participants), and correctness (\eg~technical bugs). 
After running pilots, experimenters analyze the log data or open-ended feedback along with quantitative measures, and then revisit related steps to fix (P4, P5, P14). 
Sometimes, they realize missing controls or unexpected patterns after a formal study, which becomes inspiration for a subsequent study (P15, P18). 

\subsection{Challenges: Unmet Objectives}
\begin{figure}
    \centering
    \includegraphics[alt={Two Venn diagrams illustrate the thoroughness and controllability objectives. Section A shows that the thoroughness objective aims to maximize the overlap between the target experimental space and the space covered by the stimuli. Section B shows that the controllability objective aims to minimize the difference of the stimuli set from the set under the researcher's control.},width=\linewidth]{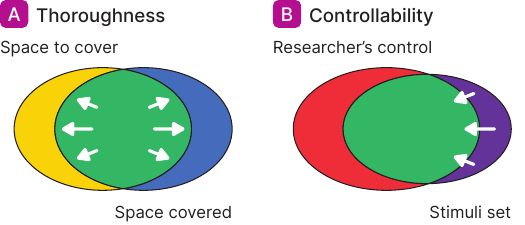}
    \caption{(A) The \emph{thoroughness} objective is to have stimuli that sufficiently cover the breadth and depth of the research space. (B) The \emph{controllability} objective is to fully govern and manage stimuli and their design details.}
    \label{fig:objectives}
\end{figure}

We characterize challenges in visualization experiment stimuli as researchers' \textit{unmet objectives} due to \textit{constraints}, as summarized in \Cref{fig:teaser}B.  
Our interviewees shared high-level motivations, or internal virtues, in designing stimuli and other parts of experiments.
Their motivations include, but are not limited to, generalizability (within a given scope), novelty (in stimuli design, analytic methods, study variables, theoretical frameworks), experimental validity, and research ethics (\eg~avoiding p-hacking).
These motivations tended to manifest as objectives regarding stimuli design.  
We categorize objectives into two groups: \textit{thoroughness} and \textit{controllability}, as illustrated in \Cref{fig:objectives}.
First, thoroughness means researchers' objectives to ensure the quality of an experiment in breadth and depth.
Thoroughness objectives include exploring a wide variety of possible stimuli ideas, having a generalizable set of stimuli, running technically sound experiment websites, and involving a broader audience by adjusting difficulty or apparatus concerns. 
Next, controllability means researchers' objectives to produce and manage stimuli as they want, while having full oversight.
For example, researchers may want to manipulate stimuli in detail to show and isolate intended effects, design custom stimuli to test novel ideas, and oversee stimuli and study platforms. 

However, there are constraints that prevent researchers from fulfilling their objectives. 
We classify those constraints in terms of amount of labor, required skills, inherent uncertainty in solutions, and (un)fixability.
First, researchers can fix some constraints, but doing so requires a lot of labor. 
For example, researchers can create and update lots of stimuli (\eg~100+), but it is time-consuming to do so.
Second, some problems require advanced skills from different areas, such as programming and statistics.
Third, researchers may eventually end up with some solutions, but they may find it difficult to justify them.
For example, they may not have clear ideas about the range of parameters for stimuli.
To address these, they may have to run multiple pilots or iterative discussions. 
Lastly, researchers just cannot solve some constraints. 
For example, the number of pilots may be constrained by financial reasons, and a researcher cannot test everything within a single study due to the limited attention of participants.

Below, we elaborate these challenges with detailed experiences of our interviewees in pursuing their objectives.
While some challenges may involve different constraints, we focus on the primary causes that interviewees identified.
For example, shipment inspection is challenging primarily because researchers often need to manually go through multiple stimuli, which better tools can support. 
On the other hand, creating sophisticated stimuli is fundamentally difficult due to limited technical skills, while also requiring some labor. 

\subsubsection{Thoroughness Objective}

We describe challenges in ensuring quality in creating and managing stimuli.

\bpstart{Labor constraint}
Shipping stimuli was a highly labor-intensive task of our interviewees.
While some checks like stratification of a random parameter or counterbalancing of conditions over time can be done in a closed form (\ie~writing a test case or checking distributions), there are other sanity checks that need manual attention, such as data collection and assignment. 
First, experimenters need to check whether stimuli are assigned to right conditions; in other words, making sure that participants are seeing what they are supposed to be seeing. 
This is different from closed form checks that are done at algorithm levels; it is more about whether a stimulus (\eg~an image file, a design specification) is correctly mapped to the study conditions.
Our interviewees tended to do it manually. 
For example, P18's lab practice was basically to go through every possible condition and see if assignments are correct.
Similarly, P1 said, ``I need to grab a paper and I need to write these conditions down and I record every single value of the data I want to collect.''
Using Qualtrics, P7 manually had to ``set up a link to an image for each stimuli corresponding to each condition'' and then ``[go] through the survey with the preview function and actually [check] if it's showing the right image.''
However, when there were too many possible conditions due to randomization, staircase design, or on-the-fly generation, interviewees could not just do 100\% but a small portion of it (P2, P5, P15, P16, P17, P19).
P16 said they only go through ``the first 10'' iterations where there were more than several thousands of possible orderings of stimuli.

This challenge persists when recycling prior stimuli that were successful due to the changes to platform. 
For example, P19 had to convert prior experiments done in different survey platforms to ReVISit for better data management. 
This recreating process took a few weeks for their team consisting of technically skilled researchers.

\bpstart{Skill constraint}
To create complete visualization stimuli beyond static images or specification-based charts, researchers may need to have advanced technical skills.
For example, P9 said that due to ``packagers and minifiers,'' it became more difficult to find source codes for real-world interactive visualizations.
P19 wanted to run a study with touch screen-specific interaction techniques on tablet devices.
However, the lack of developer tools for non-desktop-facing browsers made it highly difficult to debug those techniques, such as the live log of every user movement.

Sometimes, researchers faced unexpected technical failures, which resulted in financial and time-related consequences.
For example, P5 and P14 shared experiences where their study websites failed to collect the data. 
In P5's case, it was after adding an attention check, and they eventually had to re-run the study.
In P14's case, it was only demographic information, so they had to contact the participants with extra compensation and about 90\% of them responded back. 

\bpstart{Uncertainty constraint}
Regardless of knowing technical methods for designing stimuli, interviewees had difficulties in figuring out the scope of stimuli.
This challenge occurred across different types of stimuli creation: synthetic, real-world, and replicated designs.
First, interviewees often expanded their anecdotal real-world observations to a broader set of synthetic visualizations by looking to prior theoretical discussion to complement them (P10, P18) or generating comparative cases (P4, P14). 
In doing so, they often adopted theoretical frameworks from outside of the visualization literature (\eg~vision science, fact-checking, affordance, \etc)\footnote{As visualization is interdisciplinary, it is hard to demarcate boundaries between the visualization literature and others. Here, we simply differentiate frameworks that were \emph{originated} within visualization venues like VIS, TVCG, and EuroVis. They may still have been \emph{used} in the visualization literature.}, which required further translation jobs.
P4 and P14 noted that their chosen framework was intended for text-based communication, so it was challenging to create visualizations in a way that is ``very closely equivalent to the text-only condition'' (P14).

When using synthetic data for visualizations, interviewees adopted conventional approaches, when available, to avoid unnecessary reviewer concerns (P6, P8, P10, P17), for convenience (P8, P15), or for relationship to prior work (P9).
Yet, P8 said that the frequent (almost only) use of the Gaussian distribution for a certain task made them unsure of whether the same treatment would work with other distribution types, which was often a concern of reviewers (including themselves). 
If no convention is available or suitable for their study, interviewees relied on heuristics and tested later (P2, P15).

Next, when using real-world visualizations, interviewees struggled with collecting qualified, representative cases (P5, P14, P17), while guidelines for comprehensiveness were not straightforward.
For example, P14 struggled with ``collect[ing] [N] clean examples,'' which resulted in quite a small set of stimuli (less than 10) out of a few thousands.
But, this invited reviewer criticism regarding comprehensiveness.
To achieve better comprehensiveness (\eg~in terms of chart types), P5 and P17 had to expand their collections by adding novel designs while ensuring consistency with the existing ones.

Lastly, other interviewees wanted to compare their work with previous experiments or had questions in how prior work analyzed and interpreted responses, so they decided to run replication studies. 
P3 and P9 (separately) wanted to find better ways to explain individual differences instead of reporting averages only. 
Their goal was to replicate related prior studies as much as possible, but the prior work did not provide full replication details.
For instance, P9 mentioned that studies based on Cleveland \& McGill~\cite{cleveland1984,cleveland1986} were highly replicated in visualization experiments, yet no exact implementation of stimuli were available, which gave them uncertainty in precision.

Across stimuli types, an experiment variable can have continuous parameters (\eg~correlation, latency) or discrete levels (\eg~color hue; chart layout).
Some junior interviewees noted that finding parameters and levels often felt like ``art and not science'' (P2) because there is no standardized method or concrete guidelines. 
Even if there is a guideline, how to divide or stratify a continuous parameter was quite tricky (P2, P15).
When using real-world charts, they had to label a large volume of examples to ensure even assignment (P5, P17) and find a way to quantify certain variables like visual complexity (P6).
Because these processes often involved different, at times contradicting, opinions of collaborators, they had to eventually run a pilot study to confirm.

\bpstart{Unfixable constraint}
Our interviewees found challenges that raised important concerns, but they could not fix, such as losing comprehensiveness, participants' device types, and participants' cognitive load.
First, real-world stimuli often convey multiple factors together, causing generalizability and internal validity concerns.
P14's experiment had to use data already widely adopted in daily lives (\eg~those from government agencies or news media), and P19 wanted to involve human-made data fallacies in visualizations (\eg~improper log scale use).
Because they needed to use entire charts as is, they had to justify the lack of diverse data patterns as a limitation (P5, P14).
P5 said that expert designers tended to use bar and line charts more often than pie charts and scatter plots, so they ``dropped them and focus[ed] on line and bar,'' sacrificing comprehensiveness.
Next, interviewees were concerned with potential influences of device types other than desktops/laptops, so they had to exclude participants using mobile devices (P3, P5, P12, P13), yet P3 and P12 hoped for a method to account for device factors as an increasing number people may use crowd-sourcing platforms via smartphones. 
Lastly, when a research problem itself was quite complicated (\eg~uncertainty communication), it was not possible to use basic charts, increasing participants' cognitive load.
To balance such complexity, for example, P16 omitted some stimuli to reduce the number of trials, and P12 had to add detailed instructions and training sessions.

\subsubsection{Controllability Objective}\label{sec:control}

We report challenges in controlling and overseeing experiment stimuli.

\bpstart{Labor constraint}
Because exploration and selection stages often involve a lot of candidate stimuli, managing and overseeing them caused labor-intensive challenges.
First, management of synthetic stimuli has a version control challenge. 
For example, P2 actively used LLMs to create stimuli but found it difficult to manage AI-generated stimuli, describing the process was ``totally duct tape and glued by me in a folder and I'm constantly making copies of different code.''
They also said, ``I'm not certain which direction I'm going. So I'm constantly saving lots of different drafts and I'm constantly taking screenshots.''
P1 and P2 said that updating stimuli became even more complicated when a study condition was added or removed. 
P4 desired an easier way to propagate changes made to one stimuli to others.

In contrast to synthetic charts created by a set of parameters and metadata, experimenters need to extract such parameters from real-world charts to make sure some degree of controllability and variability in stimuli and refer back to them during analysis.
This labeling may involve a lot of manual effort (P5, P17).
Although P17 used a repository with some metadata, for example, they were not enough for operationalization.
Their team had to go through several rounds of labeling and calibration of conflicting opinions, requiring further user testing.

For internal communication and reporting purposes, interviewees wanted a way to overview the study design (\eg~stimuli laid out by variables) and export it as a figure for a paper.
For example, P15 stored 24 stimuli in a folder, which they manually opened and viewed to sanity-check; they found that this tactic was not scalable for larger experiments with 100+ stimuli.
P14 said, ``figures for papers [are] both labor intensive and a lot of designing when it's a complicated study. So, if something did that for me automatically, that would be great.''
While describing themselves as an ``AI Luddite,'' P18 noted that such automation could be ``fundamental'' to experiment studies.

\bpstart{Skill constraint}
Interviewees had skill-related difficulties in keeping design consistency, creating unconventional designs, controlling parameterization, and accounting for cross-browser experiences.
First, interviewees found it challenging to keep design consistency when they were using bespoke charts (P12), switching among tools (P7), or including non-visual stimuli like text (P2, P14, P15).
For example, P7 initially created stimuli using Excel, and later needed to switch to D3, which made it difficult to replicate some of Excel's design choices.
Second,  interviewees had challenges creating ``edge-cases'' (P3) that were not widely adopted but seemed to open up interesting opportunities (P3, P12, P16). 
P16 was trying out map designs that were ``not ... available [via] traditional visualization libraries'' like Vega-Lite~\cite{satyanarayan:vega-lite2017} or ggplot2~\cite{wickham:ggplot22010}, so had to author new JavaScript code. 

Lastly, even if experimenters exclude non-desktop devices, different monitors have different levels of resolutions and sizes, different browsers may vary in rendering the same code, and people may have personalization settings like dark mode, which interviewees learned from pilot studies (P3, P13, P15, P19).
Solving cross-browser compatibility was often beyond our interviewees' skill stacks, threatening their control over stimuli.
In response, they decided to use static images instead of on-the-fly generation, such as using Vega-Lite~\cite{satyanarayan:vega-lite2017}. 
P3 further included a screen resolution check in their study. 
In one of P13's experiments, pixel-level differences were highly important, so they had to find some workarounds to prevent zooming on browsers.
A related issue is handling of floating point numbers. 
P3 and P15 used NumPy and SciPy (on Python) to generate synthetic data points with long decimal values.
However, browsers did not render these in the desired way, which was another reason they stuck to static images.

While platforms like Qualtrics or Prolific can mitigate such technical issues, they can severely limit customizability.
For example, they do not require setting up a new server and their data collection is quite stable. 
P8, P10, and P12 found those platforms to be sufficient enough for experiments with static visualizations.
For this convenience, P10 said that they had recommended Qualtrics to masters or undergraduate students' projects.
However, such platforms were not enough to run more customized studies.
For instance, P12 needed to solicit some responses and show them in later tasks.
To do so, they had to inject custom JavaScript codes, which was challenging and added further inspection needs. 
P17 also mentioned the difficulty of customizing output formats, which added unnecessary data cleaning procedures.

\bpstart{Uncertainty constraint}
During the exploration and selection steps, a major concern, particularly for perception studies, was to find stimuli that can ``demonstrate that [intended] effects exist'' (P11), which required fine control.
P3 discussed the excitement of ``testing edge cases'' that ``[do not have] abundant literature'' but also anxiety that they ``will perhaps give you interesting results or no results,'' a sentiment that was shared by P2 as well.
More technically, our interviewees had challenges in achieving contrast and subtlety in stimuli for showing intended effects.
Some effects can be achieved via contrastive stimuli with clear differences.
For example, P4 was interested in how emojis (as in icon arrays) can invoke different affects while making decisions. 
Because some emojis have shared shapes (\eg~smiley faces), their team had to carefully find emojis that have different shapes as well as imply different emotions.
It could also mean having a clear baseline case without any treatment like text descriptions (P14).
On the other hand, there are cases where the difference among stimuli needs to be vague.
When a stimulus has an easy-to-detect data pattern (\eg~correlation of 0.95), 
people are unlikely to make wrong answers, making it difficult to compare the effectiveness of different visual design choices (P1, P15, P18).
Instead, experimenters needed to show subtle cases where other factors can lead to individual variance.
Yet, there is no simple answer for the appropriate degree of subtlety in stimuli.
For example, to see the influence of people's attitude toward certain topics on graphical perception, P15 needed to run pilots only for topic sensitivity as highly sensitive topics may overwhelm effects due to chart design.
Furthermore, P13 randomly pulled out data from a distribution, and later found that a pattern where users performed well under certain numbers (those divisible by 5), which was a confounder for task difficulty.

When using real-world stimuli, interviewees had to account for internal validity to a degree by fixing elements that interrupt with isolating main effects, such as annotations explicitly describing the chart (P9, P14), excessive visual embellishment (P5), and unfamiliar or irrelevant topics (P19).
When tasks are at a higher level than visual perception, experimenters also need to select a topic of the data (P1, P10), which may introduce additional confounders.

\bstartnc{Unfixable constraints} around controllability included the difficulty to synthesize data, real-world chart quality, and external constraints.
First, using synthetic data distributions does not mean that you can fully control them. 
For example, there may not be a way to produce synthetic data that exhibits exact data parameters such as correlation, skewness, or entropy.
So, P15 had to repeatedly create a bivariate distribution and test if it achieves a certain correlation until they filled out all the needed correlation values. 
This makes it nearly impossible to produce synthetic data on the fly, as P2 did. 
Furthermore, a targeted distribution may involve many underlying parameters with correlations or causal relationship (\eg~temporal, spatial data).
After a few months of iterations, for example, P16 learned that it was impossible to synthesize their targeted distribution for geospatial data.
At the same time, synthetic data may have additional graphical constraints like no overlapping data points (P3, P12), requiring either algorithmic or manual verification. 

Next, using real-world visualizations or data may confine the room for control.
For example, P14 noted that they had too few high-quality options once filtering out illegible charts from their corpus of real-world charts, which limited sophisticated control over stimuli.
P10 also had to carefully choose a real-world data distribution that exhibits a uniform distribution of intended parameters (pairwise differences) while not provoking any strong emotions like political topics.

Lastly, some interviewees had external constraints in choosing a certain platform or toolkit.
For example, P17 had to use a specific platform that their institution had a contract with, and P13 said that ReVISit is not compliant with local data privacy regulations (\eg~GDPR).
P10 and P4 mentioned that sometimes they needed to carefully think about the total byte count of their study materials when they actually run studies with populations with limited internet access.

\subsubsection{Merits and Challenges in using Off-the-shelf Toolkits}

Our interviewees tended to use various off-the-shelf toolkits, such as Vega-Lite~\cite{satyanarayan:vega-lite2017}, Qualtrics, and Prolific.
Among them, ReVISit~\cite{nobre2021:revisit,cutler2026:revisit} was a popular keyword across our sessions.
While only a few interviewees had a complete experience of using ReVISit to run studies, others were trying it out or had strong hopes for it. 
Overall, ReVISit had benefits including customizability (P3), more types of collectible responses (P9, P19), reusability of prior stimuli (P12), and in-situ support for stimuli inspection (P3, P13); yet some interviewees with deep skills in building study websites (P10, P12, P13) found some limitations in its capability.
For example, P19 said that its screen-recording (a response option) allowed for seeing ``where [people] click,'' which helped them to debug stimuli.
On the other hand, P10 said, ``I'm accustomed to having low-level control over my experiment design and ReVISit is this mix of expecting you to work [at a] high level, but it also expects you to have technical skills.'' 
P10 further shared concerns where adding a custom function or fixing some codes in ReVISit's source code could cause downstream problems, such as failing to properly show stimuli and collect responses.

This duality of attitudes is not unique to ReVISit. 
For example, P4 found Vega-Lite~\cite{satyanarayan:vega-lite2017} to be useful to test out different ideas by just changing some properties in a design specification, whereas P3 sometimes had difficulty using Vega-Lite to add custom interactions or control low-level visual properties like padding between elements. 
P12 shared a failure experience when adding custom JavaScript code to Qualtrics platform. 
Due to the limited customizability, their team decided to stick to custom React websites whenever their study would involve interactions.

\subsection{Strategies}\label{sec:strategies}

Interviewees shared strategies that they gained from their experiences and now use for mentoring.
We introduce some recurring, high-level patterns below, along with a list of low-level tactics in \Cref{fig:tactics}.

\bstartnc{Testing everything is a ``fool's errand''} (P10).
Emphasizing the importance of admitting there is no perfect study, interviewees desired comprehensively testing in a well-scoped space rather than ``false narrative of generalizability'' for every possible case (P10).
Specifically, P6 said such behavior would encourage P-hacking while analyzing responses, and P11 said adding too many factors would only increase more noise in responses.
P11 also mentioned that after fixing a chart type, they were able to enumerate and reason about other possible conditions in a more explainable way.
Studies comparing complete chart designs (\eg~quantile dot plot, error bars) as constructs can have fewer examples often based on popularity in literature (P4, P12) or in society (P14).
On the other hand, experiments for isolating certain perceptual variables (P2, P18) would still need to create a lot of stimuli with a couple of variables as those variables form continuous parameter spaces or have more than a few dozens of levels.

\begin{table}
    \centering
    \caption{Low-level tactics used by our interviewees. Note that these tactics are highly contextualized. The stages are: Exp(loration), Sel(lection), Ship(ment), Depl(loyment), and Anal(ysis).}
    \label{fig:tactics}
    \includegraphics[width=\linewidth]{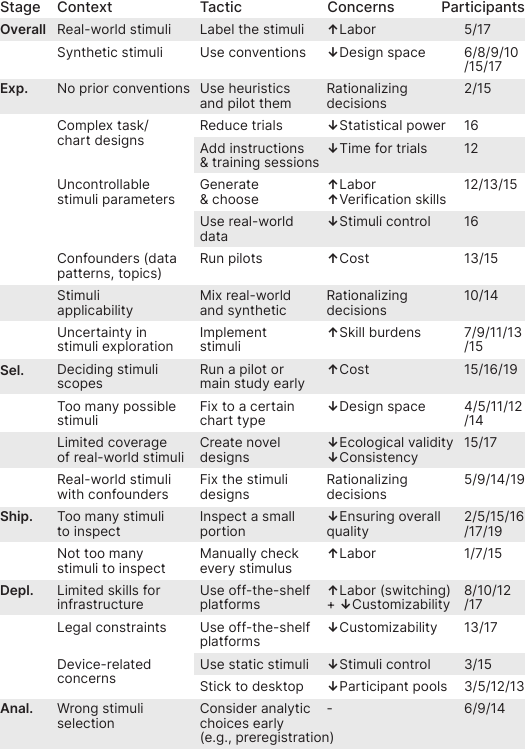}
\end{table}

\bpstart{Internal vs. ecological: Either path over middle ground}
When it comes to internal and ecological validity concerns, our interviewees tended to choose either path and mitigate remaining concerns rather than find a middle ground. 
Some used synthetic, highly controlled charts to isolate a variable of interest, admitting loss of realism (P2, P7, P14, P15, P19).
Others used real-world visualizations and data sets because they had related application domains (P5, P10, P13), wanted to convey the historical trajectories with some charts (P7, P14, P19), or were not able to synthesize one (P16).
In addition, interviewees often mixed synthetic and real-world sources. 
For example, P10 used real-world data but created synthetic charts; while P14 used real-world charts but added synthetic text elements to them.

\bpstart{\textit{Seeing vs. thinking}}
Many interviewees reported that they had gone through long periods of exploring designs.
These processes commonly include brainstorming different ideas, getting feedback from mentors and collaborators, and fixing them. 
In doing so, interviewees experienced technical burdens, such as version control and programming (P1, P2, P8, P12).
Technical complexity made P12 hesitant to explore more stimuli. 
However, implementing stimuli is an important step because there is 
``a big difference between thinking ... versus seeing ... variations'' in stimuli (P11). 
Implementation can function as a first pilot with themselves (P13, P15), produce reusable material for later stages (P7, P9), and reduce ambiguity among collaborators (P9).

\bpstart{Analysis as guidelines}
After running several experiments, P14 said, ``now I think more deeply about ... how to design an experiment in a way to make the stats later work better.''
With the same realization, P6 and P9 mentioned that they had used thinking about analysis ahead of the time as a key principle in their mentorship. 
For example, P9 said, ``when you're designing the stimuli, you might control factors better, you might structure the stimuli itself or the instrumentation around the stimuli, so that you can more readily do the types of analysis that you anticipate.''
P6 had a rule that made their students complete pre-registration and IRB, which was in turn a forcing function to reason about data analysis while designing experiments.
A challenge here is that many Ph.D. students with computer science backgrounds did not have prior statistics experiences, so they had to find related resources on their own (P2, P12) \mbox{or take courses from another department (P6, P10)}.

\bpstart{``Close the loop ASAP''}
Junior interviewees (P15 and P16) said that their advisor often instructed them to run a study based on their best understanding at the moment ``rather than planning everything'' (P15).
Once they analyzed the first study, they reasoned about what were missing or further interesting questions, which ``pave[d] the path'' (P16) for the next steps.
P9 observed that they had ``one pattern'' of ``closing the loop as soon as possible with sort of a minimal, viable experiment. ... within the lab or with a few participants at most,'' which not only made their students move on but also get some ``muscle memory'' of how to run an experiment. 
P1 also mentioned that running a small pilot study tended to be more useful because just speculating about stimuli designs often wasted their time without any confirmation. 

\subsection{Concerns and Opportunities for AI}
In our interview sessions, we asked interviewees to brainstorm how automation can help their work and what boundaries they might draw for AI-assisted visualization experiments. 
At a high level, participants voiced concerns with losing control, but they expected reduced labor and enhanced thoroughness of the stimuli scope.

\subsubsection{Concerns}

\bstartnc{``I want to be in the driver's seat.''}
Our interviewees unanimously wanted to take the control in AI-assisted experiment design, with varying degrees of openness to AI assistance. 
At the one extreme, P18 did not want any kind of automation in experiment design but coding, whereas P14, P2, and P11 had more optimism in using AI for various tasks. 
Still, they ``want[ed] to be in the driver's seat'' (P11) at every step. 
P11 was worried that, if AI does things on behalf of them, then they get ``this black box that [they] have to decode'' regarding important decisions and rationales, which could be an additional, nebulous job.
As ``someone who's very open to AI,'' P14 said, ``I wouldn't have [AI] write a section of the paper for me'' because ``it can't take accountability.''
They also pointed out that AI suggestions for confounding factors in stimuli could encourage ``fishing for hypotheses.''

Interviewees also expressed the importance of human care as researchers.
While noting that ``AI is going to continue to get better,'' P2 said that ``the care to want to solve the problem and the interest to put the time and energy'' would be something that they wanted to keep holding on. 
Plus, some interviewees did not want automation because the very process of experiment design was enjoyable.
For example, P16 said they would not use LLMs to ask about research ideas because they ``find a lot of joy in figuring out what a good study is.''

\bpstart{Bias and homogeneity}
Interviewees shared about their concerns regarding biases of AI assistance. 
For example, P10 was concerned with the poor quality of AI-based literature review and summarization, which further caused distrust in automated literature solicitation. 
More importantly, some were worried about the homogeneity of AI-suggested topics, which is already being observed in science~\cite{hao2026:focus}, and hence resulting in similar stimuli designs.
For instance, P3 pointed out historical bias of AI which could suggest stimuli designs that ``you haven't considered but other people have already [done],'' which would in turn strip the novelty of their work. 
In a similar vein, P9 noted that AI-suggested experiments would be ``more constrained than what people do now, which is almost anything.''
P14 was also worried that AI suggestions would bias them, making it difficult to ``independently verify [their] work.''
P18 was similarly concerned that their advisees would not be able to develop an ability to independently think---not only about stimuli designs, but also overall experiment designs. 

\subsubsection{Opportunities}

Below, we summarize ideas from our interviewees.
In the following section, we will discuss their potential impacts and relationship to the literature.

\bstartnc{Stimuli creation and management} was a frequently mentioned avenue for future tool support, as noted above.
This tendency was because they often found technical creation of stimuli (\eg~using D3) or setting up an experiment platform to be relatively less essential parts than experiment design and data analysis. 
P12 and P16 noted that they could explore a broader set of possible designs if they were able to promptly generate visualizations that are not well-supported by toolkits.
P12 envisioned further use cases, such as creating charts on the fly so that participants can provide more contextual qualitative feedback.
P2 and P4 emphasized the need to propagate design changes, as making consistent changes across different stimuli was challenging. 
Because LLM prompts tend to be natural language, P4 suggested a tool that generates semantically equivalent or varying icons for icon arrays.
For example, if they want to compare ``concrete'' (\eg~faces, materials) and ``abstract'' shapes (\eg~arrows), then an AI model should allow manipulations based on such concepts.
P9 suggested that automated stimuli creation could benefit ``closing the loop'' early.
Yet, such support should come with a caution.
P19 shared an experience where the stimuli and tasks were too easy for LLMs to do but slightly troublesome for people to do, resulting in LLM-generated, invalid, homogeneous responses with no error.

\bpstart{Automated shipment verification} 
Interviewees often imported stimuli images to study platforms or websites by manually matching image URLs to study conditions, which they found unscalable (P15). 
This inconvenience is further problematic when experimenters are making changes to stimuli design or conditioning after shipping it to a platform, which could mean starting from scratch (P10).
Plus, experimenters may want to reuse their stimuli from pilots partially or entirely after pilot studies, which could require another from-scratch shipment (P9).
Thus, a ``unified'' ecosystem (P15) could integrate design iteration, shipment to study platforms, verification of assignments, and reuse of stimuli (P9, P15, P16). 
As a verification method, some participants brainstormed around simulated participants, which we further discuss below. 

\bpstart{Simulated responses} 
\textbf{\textit{While all of our interviewees noted that they would never replace human responses with synthetic ones}} (formal study or parameterization pilots), they pointed to places where response simulation could be useful.
On the one hand, interviewees wanted sanity checks of their stimuli (\eg~shipment, bugs) using AI.
For example, a computer vision agent could go through the study for each possible condition and check if a stimuli is visually correct (not just by the file name) for a condition or if there are any bugs (P15, P8).
On the other hand, interviewees wished for methods to check downstream effects of their stimuli to statistical analysis (P5, P16).
For example, to assess the suitability of stimuli for different literacy levels, P5 wished for simulated responses for different participant demographics (\eg~education level).
P16 also mentioned that some realistically simulated responses could help them envision final statistical modeling and assess if their study design is internally valid. 

\bpstart{Scaffolding methods} 
Given that stimuli designs can have downstream effects (\Cref{sec:strategies}), interviewees often had difficulty reasoning about possible next steps in stimuli design, or saw students who did.
While acknowledging that many decisions do not have clear solutions, they brainstormed some potential opportunities to help with their reasoning.
For example, P2 often experienced challenges in choosing statistical methods.
Based on experiment conditions and stimuli parameter settings, they wanted an AI tool that suggests ``standard guidelines'' along with their pros and cons, so that they could reason about them and later discuss with their advisor.
Similarly, P6 imagined a linter-like tool that inspects their experiment design and helps them to understand the limitations.
Their intention was not to fix that limitations but to have confidence in what they could generalize through their experiments. 
Lastly, with P9, we brainstormed about AI-based educational tools that can instruct possible next steps based on previous decisions, acknowledging potential biases. 
For example, if a study design is tentatively decided, a simulation tool can suggest potential stimuli options and help experimenters to think about their applicability, consequences to statistical methods, \etc

%% file: sections/5-discussion.tex
\section{Discussion: Future Research Opportunities}

Based on our findings and prior literature, we outline several promising research opportunities in terms of stimuli creation and management, response simulation, cross-apparatus support, and scaffolding for experiment design.

\subsection{Stimuli Creation and Management}
Our findings confirm that one-shot generation of possible stimuli given a hypothesis will not fulfill researchers' needs at various stages of experiment design.
Instead, stimuli creation and management systems must integrate support for creating, exploring, selecting, verifying, and shipping stimuli.
Suppose that a researcher has some seed idea about stimuli and related hypotheses, and a system can enumerate a set of initial designs.
For synthetic visualizations, prior techniques in partial specification support~\cite{wongsuphasawat2016:voyager,moritz2018:formalizing} and design augmentation~\cite{kim2026:kbAug} could be useful.
For real-world visualizations, style transfer~\cite{everts2015linestyle,harper2017,Harper2014} and classification~\cite{savva2011:revision} techniques could further support filling missing chart designs based on existing ones or reduce confounders between stimuli.
Replicating interactive visualizations from real-world cases could adopt automated methods for adding interactions to static visualizations~\cite{synder2024:divi}. 
Natural language-based visualization creation~\cite{narechania2021:nl4dv,sah2024:nl4dvllm} will also play an important role for semantic consistency.
Interesting future research is to understand how AI-assistance for visualization stimuli supports or disrupts replication studies in terms of code quality and readability.

When exploring and selecting stimuli design, we point out four important requirements: propagating design changes, version control, interactions for adding/removing conditions, and addressing biases. 
First, updating stimuli consistently may not necessarily mean repeating the same changes.
For example, making changes to mark type in Vega-Lite (\eg~\texttt{point} $\rightarrow$ \texttt{rect}) can also change the color palettes (white-to-blue $\rightarrow$ Viridis). 
Plus, a common exploration trajectory in any design task is trying out an idea and going back to an original, which can happen much later in the process.
This nature requires effective version control methods (\eg~selectively canceling design changes) as well as an overview dashboard.
Third, such an overview dashboard could interactively support adding or removing conditions to adjust the study scope or match participants' cognitive capacity.
Future work can refer to responsive visualization literature~\cite{hoffswell2020:responsive,kim2023:dupo,schottler2024:breakpoints} as they share similar concerns. 
Lastly, automated stimuli creation systems should be able to assess and inform homogeneity and bias in suggested designs.
A potential method is a platform for sharing experiments without requiring technical burdens, such as refactoring their study designs.
For example, a generative AI-based method can produce a preliminary study specification from a pre-registration text (as a starting point), and the researchers can clarify it with minimum effort. 
Such a structured dataset of study designs could allow for assessing the frequency of stimuli designs and illuminating gaps that a proposed study might be able to fill.

To correctly ship multiple stimuli to a study platform, our interviewees tended to rely on tedious manual tasks. 
Thus, a stimuli management system must consider relevant automation techniques.
A potential approach is to produce stimuli in ReVISit-supported formats (\eg~React components, Vega-Lite specifications) and their assignment as a ReVISit config spec~\cite{cutler2026:revisit}.
For those with limited tool options, systems can generate equivalent configuration files for platforms like Prolific~\cite{prolificAPI} and Qualtrics~\cite{qualtricsAPI} given that they offer programmatic methods for creating a study form.
Shipment automation will require high accuracy or inspection methods, \mbox{which we discuss in the following section.}

\subsection{Guaranteed Large-scale Inspection}
Inspecting stimuli does not mean checking them one by one, but seeing them in full sequences of a given study.
For example, bugs can happen in specific orderings of stimuli, and there can be unexpected priming effects.
Furthermore, an update to stimuli after that inspection can mean additional inspection.
However, when there are more than a few dozens of stimuli, and they are ordered based on complex logic, it becomes hard to inspect them with 100\% coverage. 

We suggest two approaches---machine learning-based automated checks and finding a representative subset---for hybrid, guaranteed, large-scale inspection.
First, automated inspection systems can go through study websites, perform tasks, and check certain concerns like repeated visual similarity.
They can utilize ML-based visualization perception techniques~\cite{haehn2019:cnn,wang2025:aligned} and AI agents like Claude Cowork.  
Next, manual inspection is still essential to ensure stimuli quality.
To reduce manual efforts, algorithmic methods to find a minimum, guaranteed subset of stimuli assignments (\eg~random ordered stimuli) could be useful.
Systems could adaptively modify a guaranteed subset as inspection proceeds, based on initial automated checks.

\subsection{Response Simulation}

We clarify again that simulating experiment responses can be harmful in conducting experiments ethically. 
Our discussion about response simulation is bounded to support pre- or post-study tasks that could inform stimuli design, such as choosing statistical methods, inspecting lack of effects of stimuli before pilots, and simulating various normative scenarios given stimuli designs.
With this in mind, an interesting application scenario is mimicking different population demographics, such as literacy or education levels, as those factors can have influence on how participants interpret stimuli (\eg~P5). 
Researchers may take a closer look for cases where a performance measure drops or surges in particular demographic groups.
Inspired by a \textit{rational agent framework}~\cite{wu2024:rational} and \textit{causal quartets}~\cite{gelman2024:quartets}, responses can be simulated to represent normative statistical scenarios (\eg~no effects, too strong effects) to help researchers interpret human response data. 

There are important considerations in terms of the performance, behavior, and safeguards of automated response simulation.
First, performance of automated techniques could be problematic in simulating responses.
Prior work on ML-based visualization perception and evaluation techniques~\cite{wang2025:aligned,haehn2019:cnn,wang2025:dracogpt} tend to exhibit low alignment with human counterparts, which is also prevalent in LLM-based response generation~\cite{hullman2026:iid,grossman2023:transform}.
Next, ML-based methods may have different objectives than human-subject experiments.
Many of them tend to focus on how accurately an automated technique can process visualization images, while visualization experiments essentially want to collect how humans make errors.
A possible concern is that automated agents make no errors with stimuli designed to provoke specific human errors.
It will be an important research question whether such models should exhibit human-like errors.
For instance, while human-like behaviors could simulate patterns that are better aligned with human cognition~\cite{lake2017}, they might reinforce historical biases and harms~\cite{livesey2017}.
Lastly, we can imagine malicious researchers using response simulation techniques for data fabrication, as science history shows~\cite{fanelli2009:fabrication}.
Thus, response simulation techniques would ideally come with safeguards, such as watermarking (\cf~\cite{pan2024:markllm,alMarzouki2005:real}) that is robust against data removal or permutation.

\subsection{Accounting for Apparatus}
As described in \Cref{sec:control}, our interviewees had controllability challenges regarding apparatus, stemming from the fact that they cannot fully anticipate and control participant devices in crowd-sourcing settings. 
In practice, browsers may behave differently on desktop and smartphones, and mobile devices have a wide range of screen sizes, which makes it nearly impossible for even professional designers to inspect them all~\cite{hoffswell2020:responsive}.
Due to discrepancies in browser behaviors, interviewees could not test stimuli design on all possible platforms even if they wanted to recruit participants on diverse devices. 
Furthermore, it is extremely difficult to obtain real-time logs for interactions with stimuli (as we do with browser developer tools) on smartphones and tablets, as P19 indicated, so debugging capacity is limited.
To mitigate this technical difficulty, responsive visualization-based approaches could be useful with careful coordination of perceptual differences with respect to physical devices (\eg~same pixels may result in different millimeters).
While tools like Dupo~\cite{kim2023:dupo} propose some physical size simulation methods, every device has different pixel resolutions, so a more granular control is necessary, ideally using perceptual units like a degree of visual angle.
At the same time, future work is warranted to extend off-the-shelf toolkits and study platforms to support robust cross-device experiences.

\subsection{Scaffolding}

Stimuli creation and management cannot be separated from the overall experiment design processes. 
For learning purposes, some interviewees brainstormed scaffolding support in experiment design. 
For example, given a hypothesis, an educational system could provide possible  applicable next steps, such as nudging how many stimuli are needed in what study conditions, instead of suggesting the results directly.
Potentially, an early analysis of participants' time and cognitive load would help assess the study scope.
Based on stimuli design, their assignment, and simulated responses, such a system could further inform analytic choices, adopting modeling support systems~\cite{tea2019:tea,jun2022:tisane,jun2024:rtisane,guo2025:vmc} and multiverse analysis systems~\cite{liu2021:boba,sarma2023:multiverse,saram2024:milliways}, for example.
With P9 we brainstormed visual demonstrations of influences of previous decisions to next steps, inspired by \textit{Seeing Theory}~\cite{seeingTheory} that illustrates how statistical distributions and testing methods work in a detailed step-by-step animation, as a method to amplify the educational value of scaffolding systems for experiment design.

%% file: sections/6-conclusion.tex
\section{Conclusion}

Stimuli play an essential role in visualization experiments and need tremendous care of researchers.
Yet, researchers strategies, challenges, and needs regarding stimuli management have remained untold. 
To understand practices and illuminate future paths, we conducted a semi-structured interview study with 19 visualization experimenters with diverse levels of experience from 14 different institutions. 
Our findings inform how researchers' motivations in creating and managing stimuli thoroughly with controllability are constrained due to labor, skill, uncertainty, and (un)fixability, as well as their strategies. 
We also report optimism in AI for facilitating such processes as well as concerns and criticism regarding the use of AI for visualization science. 
Based on our findings, we discuss future research opportunities in terms of stimuli creation and management, guaranteed shipment verification, response simulation, and apparatus concerns. 

\bpstart{Limitations}
Our work has multiple limitations with respect to participant demographics.
Our participants tended to have experiences in online, crowd-sourcing experiments, reflecting its wide use in the field.
A large portion of participants were junior and early-career researchers as we prioritized capturing recent, hands-on experiences.
Lastly, except for a few participants who were interested in individual variances, most of them designed their experiments based on a positivist paradigm. 
Future work can extend our work regarding the above aspects.

%% file: reference.bib
@article{lam2012:scenarios,
	title        = {Empirical Studies in Information Visualization: Seven Scenarios},
	author       = {Lam, Heidi and Bertini, Enrico and Isenberg, Petra and Plaisant, Catherine and Carpendale, Sheelagh},
	year         = 2012,
	journal      = {IEEE Trans. Vis. Comput. Graphics},
	volume       = 18,
	number       = 9,
	pages        = {1520--1536},
	doi          = {10/drrh6j}
}

@article{elliot2021:methods,
	title        = {A Design Space of Vision Science Methods for Visualization Research},
	author       = {Elliott, Madison A. and Nothelfer, Christine and Xiong, Cindy and Szafir, Danielle Albers},
	year         = 2021,
	journal      = {IEEE Trans. Vis. Comput. Graphics},
	volume       = 27,
	number       = 2,
	pages        = {1117--1127},
	doi          = {10/ghgt5q}
}

@inproceedings{nobre2024:literacy,
	title        = {Reading Between the Pixels: Investigating the Barriers to Visualization Literacy},
	author       = {Nobre, Carolina and Zhu, Kehang and M\"{o}rth, Eric and Pfister, Hanspeter and Beyer, Johanna},
	year         = 2024,
	booktitle    = {ACM Human Factors in Computing Systems},
	publisher    = {ACM},
	series       = {CHI '24},
	doi          = {10/n98g},
    pages        = {197:1--17}
}

@inbook{ziemkiewicz2020:open,
	title        = {Open Challenges in Empirical Visualization Research},
	author       = {Ziemkiewicz, Caroline and Chen, Min and Laidlaw, David H. and Preim, Bernhard and Weiskopf, Daniel},
	year         = 2020,
	booktitle    = {Foundations of Data Visualization},
	publisher    = {Springer International Publishing},
	address      = {Cham},
	pages        = {243--252},
	doi          = {10/qw5h},
	editor       = {Chen, Min and Hauser, Helwig and Rheingans, Penny and Scheuermann, Gerik}
}

@inproceedings{plaisant2004:challenge,
	title        = {The challenge of information visualization evaluation},
	author       = {Plaisant, Catherine},
	year         = 2004,
	booktitle    = {Proceedings of the Working Conference on Advanced Visual Interfaces},
	publisher    = {ACM},
	series       = {AVI '04},
	pages        = {109--116},
	doi          = {10/c7g3x4}
}

@inbook{carpendale2008:eval,
	title        = {Evaluating Information Visualizations},
	author       = {Carpendale, Sheelagh},
	year         = 2008,
	booktitle    = {Information Visualization: Human-Centered Issues and Perspectives},
	publisher    = {Springer-Verlag},
	address      = {Berlin, Heidelberg},
	pages        = {19--45}
}

@article{borgo2018:crowd,
	title        = {Information Visualization Evaluation Using Crowdsourcing},
	author       = {Borgo, R. and Micallef, L. and Bach, B. and McGee, F. and Lee, B.},
	year         = 2018,
	journal      = {Comput. Graphics Forum},
	volume       = 37,
	number       = 3,
	pages        = {573--595},
	doi          = {10/gdxdk8}
}

@inproceedings{burns2020:level,
	title        = {How to evaluate data visualizations across different levels of understanding},
	author       = {Burns, Alyxander and Xiong, Cindy and Franconeri, Steven and Cairo, Alberto and Mahyar, Narges},
	year         = 2020,
	booktitle    = {2020 IEEE Workshop on Evaluation and Beyond - Methodological Approaches to Visualization (BELIV)},
	pages        = {19--28},
	doi          = {10/kz34}
}

@article{cutler2026:revisit,
	title        = {{ReVISit 2}: A Full Experiment Life Cycle User Study Framework},
	author       = {Cutler, Zach and Wilburn, Jack and Shrestha, Hilson and Ding, Yiren and Bollen, Brian and Nadib, Khandaker Abrar and He, Tingying and McNutt, Andrew and Harrison, Lane and Lex, Alexander},
	year         = 2026,
	journal      = {IEEE Trans. Vis. Comput. Graphics},
	volume       = 32,
	number       = 1,
	pages        = {13--23},
	doi          = {10/hbkxwp}
}

@inproceedings{nobre2021:revisit,
	title        = {{ReVISit}: Looking Under the Hood of Interactive Visualization Studies},
	author       = {Nobre, Carolina and Wootton, Dylan and Cutler, Zach and Harrison, Lane and Pfister, Hanspeter and Lex, Alexander},
	year         = 2021,
	booktitle    = {ACM Human Factors in Computing Systems},
	publisher    = {ACM},
	series       = {CHI '21},
	doi          = {10/gksk5b},
	pages        = {25:1--13}
}

@article{okoe2015:grapunit,
	title        = {{GraphUnit}: Evaluating Interactive Graph Visualizations Using Crowdsourcing},
	author       = {Okoe, Mershack and Jianu, Radu},
	year         = 2015,
	journal      = {Comput. Graphics Forum},
	volume       = 34,
	number       = 3,
	pages        = {451--460},
	doi          = {10/f7kr8h}
}

@inproceedings{jianu2025:visunit,
	title        = {{VisUnit}: Literate Visualisation Studies Assembled from Reusable Test-Suites},
	author       = {Jianu, Radu and Slingsby, Aidan and Laksono, Dany and Okoe, Mershack},
	year         = 2025,
	booktitle    = {ACM Human Factors in Computing Systems},
	publisher    = {ACM},
	series       = {CHI '25},
	doi          = {10/qw5j},
	pages        = {917:1--15}
}

@article{stoet2017:psytoolkit,
	title        = {{PsyToolkit}: A Novel Web-Based Method for Running Online Questionnaires and Reaction-Time Experiments},
	author       = {Gijsbert Stoet},
	year         = 2017,
	journal      = {Teaching of Psychology},
	volume       = 44,
	number       = 1,
	pages        = {24--31},
	doi          = {10/gftft5}
}

@article{deleeuw2023:jspsych,
	title        = {{jsPsych}: Enabling an Open-Source Collaborative Ecosystem of Behavioral Experiments},
	author       = {de Leeuw, Joshua R. and Gilbert, Rebecca A. and Luchterhandt, Björn},
	year         = 2023,
	journal      = {Journal of Open Source Software},
	publisher    = {The Open Journal},
	volume       = 8,
	number       = 85,
	pages        = {5351:1--4},
	doi          = {10/gtzcxz}
}

@article{peirce2007:psychopy,
	title        = {{PsychoPy}—Psychophysics software in Python},
	author       = {Jonathan W. Peirce},
	year         = 2007,
	journal      = {J. Neurosci. Methods},
	volume       = 162,
	number       = 1,
	pages        = {8--13},
	doi          = {10/b5dqq8}
}

@article{peirce2019:psychopy2,
	title        = {{PsychoPy2}: Experiments in behavior made easy},
	author       = {Peirce, Jonathan and Gray, Jeremy R. and Simpson, Sol and MacAskill, Michael and H{\"o}chenberger, Richard and Sogo, Hiroyuki and Kastman, Erik and Lindel{\o}v, Jonas Kristoffer},
	year         = 2019,
	journal      = {Behavior Research Methods},
	volume       = 51,
	number       = 1,
	pages        = {195--203},
    doi          = {10/gft89w},
	doiOiriginal = {10.3758/s13428-018-01193-y}
}

@inproceedings{jun2022:tisane,
	title        = {Tisane: Authoring Statistical Models via Formal Reasoning from Conceptual and Data Relationships},
	author       = {Jun, Eunice and Seo, Audrey and Heer, Jeffrey and Just, Ren\'{e}},
	year         = 2022,
	booktitle    = {ACM Human Factors in Computing Systems},
	publisher    = {ACM},
	series       = {CHI '22},
    doi          = {10/hbsvwh},
	doiOiriginal = {10.1145/3491102.3501888},
	pages        = {490:1--16}
}

@inproceedings{tea2019:tea,
	title        = {Tea: A High-level Language and Runtime System for Automating Statistical Analysis},
	author       = {Jun, Eunice and Daum, Maureen and Roesch, Jared and Chasins, Sarah and Berger, Emery and Just, Rene and Reinecke, Katharina},
	year         = 2019,
	booktitle    = {ACM Symposium on User Interface Software and Technology},
	publisher    = {ACM},
	series       = {UIST '19},
	pages        = {591--603},
	doi          = {10/gmbdfj},
    doiOiriginal = {10.1145/3332165.3347940}
}

@inproceedings{jun2024:rtisane,
	title        = {{rTisane}: Externalizing conceptual models for data analysis prompts reconsideration of domain assumptions and facilitates statistical modeling},
	author       = {Jun, Eunice and Misback, Edward and Heer, Jeffrey and Just, Rene},
	year         = 2024,
	booktitle    = {ACM Human Factors in Computing Systems},
	publisher    = {ACM},
	series       = {CHI '24},
	doi          = {10/qw5k},
    doiOiriginal = {10.1145/3613904.3642267},
	pages        = {1037:1--16}
}

@inproceedings{lam2024:lloom,
	title        = {Concept Induction: Analyzing Unstructured Text with High-Level Concepts Using {LLooM}},
	author       = {Lam, Michelle S. and Teoh, Janice and Landay, James A. and Heer, Jeffrey and Bernstein, Michael S.},
	year         = 2024,
	booktitle    = {ACM Human Factors in Computing Systems},
	publisher    = {ACM},
	series       = {CHI '24},
	doi          = {10/qw5m},
    doiOiriginal = {10.1145/3613904.3642830},
	pages        = 766
}

@article{musslick2025:aiScience,
	title        = {Automating the practice of science: Opportunities, challenges, and implications},
	author       = {Sebastian Musslick  and Laura K. Bartlett  and Suyog H. Chandramouli  and Marina Dubova  and Fernand Gobet  and Thomas L. Griffiths  and Jessica Hullman  and Ross D. King  and J. Nathan Kutz  and Christopher G. Lucas  and Suhas Mahesh  and Franco Pestilli  and Sabina J. Sloman  and William R. Holmes},
	year         = 2025,
	journal      = {Proceedings of the National Academy of Sciences},
	volume       = 122,
	number       = 5,
	pages        = {e2401238121:1--11},
	doi          = {10/g84s7d},
    doiOiriginal = {10.1073/pnas.2401238121}
}

@article{bianchini2022:aiScience,
	title        = {Artificial intelligence in science: An emerging general method of invention},
	author       = {Stefano Bianchini and Moritz Müller and Pierre Pelletier},
	year         = 2022,
	journal      = {Research Policy},
	volume       = 51,
	number       = 10,
	pages        = {104604:1--15},
	doi          = {10/gqm63c},
    doiOiriginal = {10.1016/j.respol.2022.104604}
}

@misc{an2024:vitality2,
	title        = {{vitaLITy 2}: Reviewing Academic Literature Using Large Language Models},
	author       = {Hongye An and Arpit Narechania and Emily Wall and Kai Xu},
	year         = 2024,
    note         = {NLVIZ Workshop at IEEE VIS '24. \url{https://arxiv.org/abs/2408.13450}}
}

@article{arpit2022:vitality,
	title        = {{VITALITY}: Promoting Serendipitous Discovery of Academic Literature with Transformers \& Visual Analytics},
	author       = {Narechania, Arpit and Karduni, Alireza and Wesslen, Ryan and Wall, Emily},
	year         = 2022,
	journal      = {IEEE Trans. Vis. Comput. Graphics},
	volume       = 28,
	number       = 1,
	pages        = {486--496},
	doi          = {10/g6sc9c},
    doiOiriginal = {10.1109/TVCG.2021.3114820}
}

@inproceedings{choe2024:papers,
	title        = {Supporting Novice Researchers to Write Literature Review using Language Models},
	author       = {Choe, Kiroong and Park, Seokhyeon and Jung, Seokweon and Kim, Hyeok and Yang, Ji Won and Hong, Hwajung and Seo, Jinwook},
	year         = 2024,
	booktitle    = {ACM Human Factors in Computing Systems},
	publisher    = {ACM},
	series       = {CHI EA '24},
	doi          = {10/qw5n},
    doiOiriginal = {10.1145/3613905.3650787},
    pages        = {307:1--9}
}

@article{agarwal2025:litllms,
	title        = {Lit{LLM}s, {LLM}s for Literature Review: Are we there yet?},
	author       = {Shubham Agarwal and Gaurav Sahu and Abhay Puri and Issam H. Laradji and Krishnamurthy Dj Dvijotham and Jason Stanley and Laurent Charlin and Christopher Pal},
	year         = 2025,
	journal      = {Transactions on Machine Learning Research},
	issn         = {2835-8856},
	url          = {https://openreview.net/forum?id=heeJqQXKg7},
}

@article{merchang2023:materials,
	title        = {Scaling deep learning for materials discovery},
	author       = {Merchant, Amil and Batzner, Simon and Schoenholz, Samuel S. and Aykol, Muratahan and Cheon, Gowoon and Cubuk, Ekin Dogus},
	year         = 2023,
	journal      = {Nature},
	volume       = 624,
	number       = 7990,
	pages        = {80--85},
	doi          = {10/gs7sbc},
    doiOiriginal = {10.1038/s41586-023-06735-9}
}

@article{grossman2023:transform,
	title        = {{AI} and the transformation of social science research},
	author       = {Igor Grossmann  and Matthew Feinberg  and Dawn C. Parker  and Nicholas A. Christakis  and Philip E. Tetlock  and William A. Cunningham},
	year         = 2023,
	journal      = {Science},
	volume       = 380,
	number       = 6650,
	pages        = {1108--1109},
	doi          = {10/gscnp5},
    doiOiriginal = {10.1126/science.adi1778}
}

@article{vanNoorden2023:survey,
	title        = {{AI} and science: what 1,600 researchers think},
	author       = {Van Noorden, Richard and Perkel, Jeffrey M.},
	year         = 2023,
	journal      = {Nature},
	volume       = 621,
	pages        = {672--675},
	doi          = {10/gtw8f2},
    doiOiriginal = {10.1038/d41586-023-02980-0},
}

@article{hao2026:focus,
	title        = {Artificial intelligence tools expand scientists'impact but contract science's focus},
	author       = {Hao, Qianyue and Xu, Fengli and Li, Yong and Evans, James},
	year         = 2026,
	journal      = {Nature},
	volume       = 649,
	number       = 8099,
	pages        = {1237--1243},
	doi          = {10/hbj27r},
    doiOiriginal = {10.1038/s41586-025-09922-y}
}

@misc{hullman2026:iid,
	title        = {This human study did not involve human subjects: Validating {LLM} simulations as behavioral evidence},
	author       = {Hullman, Jessica and Broska, David and Sun Huaman and Shaw Aaron},
	year         = 2025,
	doi          = {10/qw5p},
    doiOiriginal = {10.48550/arXiv.2602.15785},
	note         = {Working manuscript}
}

@inproceedings{zeng2023:dataset,
	title        = {A Review and Collation of Graphical Perception Knowledge for Visualization Recommendation},
	author       = {Zeng, Zehua and Battle, Leilani},
	year         = 2023,
	booktitle    = {ACM Human Factors in Computing Systems},
	publisher    = {ACM},
	series       = {CHI '23},
    pages        = {820:1--16},
	doi          = {10/kz36},
    doiOiriginal = {10.1145/3544548.3581349}
}

@article{zeng2024:tooManyCooks,
	title        = {Too Many Cooks: Exploring How Graphical Perception Studies Influence Visualization Recommendations in Draco},
	author       = {Zeng, Zehua and Yang, Junran and Moritz, Dominik and Heer, Jeffrey and Battle, Leilani},
	year         = 2024,
	journal      = {IEEE Trans. Vis. Comput. Graph.},
	publisher    = {IEEE},
	volume       = 30,
	number       = 1,
	pages        = {1063--1073},
	doi          = {10/qw5q},
    doiOiriginal = {10.1109/TVCG.2023.3326527}
}

@article{moritz2018:formalizing,
	title        = {Formalizing visualization design knowledge as constraints: Actionable and extensible models in {Draco}},
	author       = {Moritz, Dominik and Wang, Chenglong and Nelson, Greg L and Lin, Halden and Smith, Adam M and Howe, Bill and Heer, Jeffrey},
	year         = 2018,
	journal      = {IEEE Trans. Vis. Comput. Graph.},
	publisher    = {IEEE},
	volume       = 25,
	number       = 1,
	pages        = {438--448},
	doi          = {10/cs68},
    doiOiriginal = {10.1109/TVCG.2018.2865240}
}

@inproceedings{yang2023:draco2,
	title        = {{Draco} 2: An Extensible Platform to Model Visualization Design},
	author       = {Yang, Junran and Gyarmati, P\'{e}ter Ferenc and Zeng, Zehua and Moritz, Dominik},
	year         = 2023,
	booktitle    = {IEEE Visualization and Visual Analytics},
	publisher    = {IEEE},
	series       = {VIS '23},
	pages        = {166--170},
	doi          = {10/qw5r},
    doiOiriginal = {10.1109/VIS54172.2023.00042}
}

@article{wongsuphasawat2016:voyager,
	title        = {Voyager: Exploratory Analysis via Faceted Browsing of Visualization Recommendations},
	author       = {Wongsuphasawat, Kanit and Moritz, Dominik and Anand, Anushka and Mackinlay, Jock and Howe, Bill and Heer, Jeffrey},
	year         = 2016,
	journal      = {IEEE Trans. Vis. Comput. Graph.},
	volume       = 22,
	number       = 1,
	pages        = {649--658},
	doi          = {10/bdsz},
    doiOiriginal = {10.1109/TVCG.2015.2467191}
}

@inproceedings{wongsuphasawat2017:voyager,
	title        = {Voyager 2: Augmenting visual analysis with partial view specifications},
	author       = {Wongsuphasawat, Kanit and Qu, Zening and Moritz, Dominik and Chang, Riley and Ouk, Felix and Anand, Anushka and Mackinlay, Jock and Howe, Bill and Heer, Jeffrey},
	year         = 2017,
	journal      = {ACM Human Factors in Computing Systems},
	booktitle    = {Proc. CHI},
	publisher    = {ACM},
	series       = {CHI '17},
	pages        = {2648--2659},
	doi          = {10/b7jv},
    doiOiriginal = {10.1145/3025453.3025768}
}

@article{kim2023:dupo,
	title        = {Dupo: A Mixed-initiative Authoring Tool for Responsive Visualization},
	author       = {Kim, Hyeok and Rossi, Ryan and Hullman, Jessica and Hoffswell, Jane},
	year         = 2024,
	journal      = {IEEE Trans. Vis. Comput. Graph.},
	volume       = 30,
	number       = 1,
	pages        = {934--943},
	doi          = {10/qw5s},
    doiOiriginal = {10.1109/TVCG.2023.3326583}
}

@inproceedings{hu19:vizml,
	title        = {{VizML}: A Machine Learning Approach to Visualization Recommendation},
	author       = {Hu, Kevin and Bakker, Michiel A. and Li, Stephen and Kraska, Tim and Hidalgo, C\'{e}sar},
	year         = 2019,
	booktitle    = {ACM Human Factors in Computing Systems},
	publisher    = {ACM},
	series       = {CHI '19},
	pages        = {128:1--12},
	doi          = {10/gf2b86},
    doiOiriginal = {10.1145/3290605.3300358}
}

@article{pandey2023:genorec,
	title        = {{GenoREC}: A Recommendation System for Interactive Genomics Data Visualization},
	author       = {Pandey, Aditeya and L'Yi, Sehi and Wang, Qianwen and Borkin, Michelle A. and Gehlenborg, Nils},
	year         = 2023,
	journal      = {IEEE Trans. Vis. Comput. Graph.},
	volume       = 29,
	number       = 1,
	pages        = {570--580},
	doi          = {10/qw5t},
    doiOiriginal = {10.1109/TVCG.2022.3209407}
}

@article{zhang2024:adavis,
	title        = {{AdaVis}: Adaptive and Explainable Visualization Recommendation for Tabular Data},
	author       = {Zhang, Songheng and Li, Haotian and Qu, Huamin and Wang, Yong},
	year         = 2024,
	journal      = {IEEE Trans. Vis. Comput. Graph.},
	volume       = 30,
	number       = 9,
	pages        = {5923--5938},
	doi          = {10/g94bsg},
    doiOiriginal = {10.1109/TVCG.2023.3316469}
}

@article{li2022:kg4vis,
	title        = {{KG4Vis}: A Knowledge Graph-Based Approach for Visualization Recommendation},
	author       = {Li, Haotian and Wang, Yong and Zhang, Songheng and Song, Yangqiu and Qu, Huamin},
	year         = 2022,
	journal      = {IEEE Trans. Vis. Comput. Graph.},
	volume       = 28,
	number       = 1,
	pages        = {195--205},
	doi          = {10/gndq6h},
    doiOiriginal = {10.1109/TVCG.2021.3114863}
}

@article{pandey2023:medley,
	title        = {{MEDLEY}: Intent-based Recommendations to Support Dashboard Composition},
	author       = {Pandey, Aditeya and Srinivasan, Arjun and Setlur, Vidya},
	year         = 2023,
	journal      = {IEEE Trans. Vis. Comput. Graph.},
	volume       = 29,
	number       = 1,
	pages        = {1135--1145},
	doi          = {10/g9g7qk},
    doiOiriginal = {10.1109/TVCG.2022.3209421}
}

@article{wu2022:multivision,
	title        = {{MultiVision}: Designing Analytical Dashboards with Deep Learning Based Recommendation},
	author       = {Wu, Aoyu and Wang, Yun and Zhou, Mengyu and He, Xinyi and Zhang, Haidong and Qu, Huamin and Zhang, Dongmei},
	year         = 2022,
	journal      = {IEEE Trans. Vis. Comput. Graph.},
	volume       = 28,
	number       = 1,
	pages        = {162--172},
	doi          = {10/gm393p},
    doiOiriginal = {10.1109/TVCG.2021.3114826}
}

@article{haehn2019:cnn,
	title        = {Evaluating `Graphical Perception' with {CNN}s},
	author       = {Haehn, Daniel and Tompkin, James and Pfister, Hanspeter},
	year         = 2019,
	journal      = {IEEE Trans. Vis. Comput. Graphics},
	volume       = 25,
	number       = 1,
	pages        = {641--650},
	doi          = {10/gd52dt},
    doiOiriginal = {10.1109/TVCG.2018.2865138}
}

@article{wang2025:dracogpt,
	title        = {{DracoGPT}: Extracting Visualization Design Preferences from Large Language Models},
	author       = {Wang, Huichen Will and Gordon, Mitchell and Battle, Leilani and Heer, Jeffrey},
	year         = 2025,
	journal      = {IEEE Trans. Vis. Comput. Graph.},
	volume       = 31,
	number       = 1,
	pages        = {710--720},
	doi          = {10/qw5w},
    doiOiriginal = {10.1109/TVCG.2024.3456350}
}

@article{kim2026:kbAug,
	title        = {Data Augmentation for Visualization Design Knowledge Bases},
	author       = {Kim, Hyeok and Heer, Jeffrey},
	year         = 2026,
	journal      = {IEEE Trans. Vis. Comput. Graphics},
	volume       = 32,
	number       = 1,
	pages        = {538--548},
	doi          = {10/qw5x},
    doiOiriginal = {10.1109/TVCG.2025.3634811}
}

@article{wu2022:ai4vis,
	title        = {{AI4VIS}: Survey on Artificial Intelligence Approaches for Data Visualization},
	author       = {Wu, Aoyu and Wang, Yun and Shu, Xinhuan and Moritz, Dominik and Cui, Weiwei and Zhang, Haidong and Zhang, Dongmei and Qu, Huamin},
	year         = 2022,
	journal      = {IEEE Trans. Vis. Comput. Graphics},
	volume       = 28,
	number       = 12,
	pages        = {5049--5070},
	doi          = {10/grxn3d},
    doiOiriginal = {10.1109/TVCG.2021.3099002}
}

@article{cleveland1986,
	title        = {An experiment in graphical perception},
	author       = {William S. Cleveland and Robert McGill},
	year         = 1986,
	journal      = {International Journal of Man-Machine Studies},
	volume       = 25,
	number       = 5,
	pages        = {491--500},
	doi          = {10/b6vt7v},
    doiOiriginal = {10.1016/S0020-7373(86)80019-0}
}

@article{cleveland1984,
	title        = {Graphical Perception: Theory, Experimentation, and Application to the Development of Graphical Methods},
	author       = {William S. Cleveland and Robert McGill},
	year         = 1984,
	journal      = {J. Am. Stat. Assoc},
	publisher    = {Taylor \& Francis},
	volume       = 79,
	number       = 387,
	pages        = {531--554},
	doi          = {10/gdvmwd},
    doiOiriginal = {10.1080/01621459.1984.10478080}
}

@article{wickham:ggplot22010,
	title        = {A Layered Grammar of Graphics},
	author       = {Hadley Wickham},
	year         = 2010,
	journal      = {J. Comput. Graphical Stat.},
	publisher    = {Taylor \& Francis},
	volume       = 19,
	number       = 1,
	pages        = {3--28},
	doi          = {10/cttwkf},
    doiOiriginal = {10.1198/jcgs.2009.07098}
}

@article{satyanarayan:vega-lite2017,
	title        = {{Vega-Lite}: A Grammar of Interactive Graphics},
	author       = {Arvind Satyanarayan AND Dominik Moritz AND Kanit Wongsuphasawat AND Jeffrey Heer},
	year         = 2017,
	journal      = {IEEE Trans. Visualization \& Comp. Graphics},
    volume       = 23,
	number       = 1,
	pages        = {341--350},
	doi          = {10/f92f32},
    doiOiriginal = {10.1109/TVCG.2016.2599030}
}

@inproceedings{kim2017:graphscape,
	title        = {{GraphScape}: A Model for Automated Reasoning about Visualization Similarity and Sequencing},
	author       = {Kim, Younghoon AND Wongsuphasawat, Kanit AND Hullman, Jessica AND Heer, Jeffrey},
	year         = 2017,
	booktitle    = {ACM Human Factors in Computing Systems},
    series       = {CHI '17},
    pages        = {2628--2638},
	doi          = {10/gd7hg5},
    doiOiriginal = {10.1145/3025453.3025866}
}

@inproceedings{liu2018:colormap,
	title        = {Somewhere Over the Rainbow: An Empirical Assessment of Quantitative Colormaps},
	author       = {Liu, Yang and Heer, Jeffrey},
	year         = 2018,
	booktitle    = {ACM Human Factors in Computing Systems},
	publisher    = {ACM},
	series       = {CHI '18},
	pages        = {598:1--12},
	doi          = {10/gfxbcb},
    doiOiriginal = {10.1145/3173574.3174172}
}

@article{bujack2018:colormap,
	title        = {The Good, the Bad, and the Ugly: A Theoretical Framework for the Assessment of Continuous Colormaps},
	author       = {Bujack, Roxana and Turton, Terece L. and Samsel, Francesca and Ware, Colin and Rogers, David H. and Ahrens, James},
	year         = 2018,
	journal      = {IEEE Trans. Vis. Comput. Graphics},
	volume       = 24,
	number       = 1,
	pages        = {923--933},
	doi          = {10/gcqb9w},
    doiOiriginal = {10.1109/TVCG.2017.2743978}
}

@inproceedings{Harper2014,
	title        = {Deconstructing and Restyling {D3} Visualizations},
	author       = {Harper, Jonathan and Agrawala, Maneesh},
	year         = 2014,
	booktitle    = {ACM Symposium on User Interface Software and Technology},
	publisher    = {ACM},
	series       = {UIST '14},
	doi          = {10/ghvkfk},
    doiOiriginal = {10.1145/2642918.2647411},
	pages        = {253--262}
}

@article{harper2017,
	title        = {Converting basic {D3} charts into reusable style templates},
	author       = {Harper, Jonathan and Agrawala, Maneesh},
	year         = 2017,
	journal      = {IEEE Trans. Vis. Comput. Graphics},
    volume       = 24,
	number       = 1,
	pages        = {1274--1286},
	doi          = {10/gcxtgt},
    doiOiriginal = {10.1109/TVCG.2017.2659744}
}

@misc{everts2015linestyle,
	title        = {Interactive Illustrative Line Styles and Line Style Transfer Functions for Flow Visualization},
	author       = {Maarten H. Everts and Henk Bekker and Jos B. T. M. Roerdink and Tobias Isenberg},
	year         = 2015,
	url          = {https://arxiv.org/abs/1503.05787}
}

@article{narechania2021:nl4dv,
	title        = {{NL4DV}: A Toolkit for Generating Analytic Specifications for Data Visualization from Natural Language Queries},
	author       = {Narechania, Arpit and Srinivasan, Arjun and Stasko, John},
	year         = 2021,
	journal      = {IEEE Trans. Vis. Comput. Graphics},
	volume       = 27,
	number       = 2,
	pages        = {369--379},
	doi          = {10/ghgthp},
    doiOiriginal = {10.1109/TVCG.2020.3030378}
}

@misc{sah2024:nl4dvllm,
    title={Generating Analytic Specifications for Data Visualization from Natural Language Queries using Large Language Models}, 
    author={Subham Sah and Rishab Mitra and Arpit Narechania and Alex Endert and John Stasko and Wenwen Dou},
    year={2024},
    note =  {NLVIZ Workshop at IEEE VIS '24. \url{https://arxiv.org/abs/2408.13391}},
}

@article{schottler2024:breakpoints,
	title        = {Constraint-Based Breakpoints for Responsive Visualization Design and Development},
	author       = {Sch\"{o}ttler, Sarah and Dykes, Jason and Wood, Jo and Hinrichs, Uta and Bach, Benjamin},
	year         = 2024,
	journal      = {IEEE Trans. Vis. Comput. Graphics},
	pages        = {4593--4604},
	doi          = {10/m3f3},
    doiOiriginal = {10.1109/TVCG.2024.3410097}
}

@inproceedings{hoffswell2020:responsive,
	title        = {Techniques for Flexible Responsive Visualization Design},
	author       = {Jane Hoffswell AND Wilmot Li AND Zhicheng Liu},
	year         = 2020,
	booktitle    = {ACM Human Factors in Computing Systems},
	publisher    = {ACM},
	series       = {CHI '20},
	pages        = {1--13},
	doi          = {10/gk9mqp},
    doiOiriginal = {10.1145/3313831.3376777}
}

@misc{prolificAPI,
	title        = {Your first data collection},
	author       = {Prolific},
	year         = 2026,
	note         = {Last visited  Mar. 18, 2026. \url{https://docs.prolific.com/documentation/get-started/your-first-data-collection}}
}

@misc{qualtricsAPI,
	title        = {{Survey API Introduction}},
	author       = {Qualtrics},
	year         = 2026,
	note         = {Last visited  Mar. 18, 2026. \url{https://api.qualtrics.com/f8fdcc4586e1f-survey-api-introduction}}
}

@article{wang2025:aligned,
	title        = {How Aligned are Human Chart Takeaways and LLM Predictions? A Case Study on Bar Charts with Varying Layouts},
	author       = {Wang, Huichen Will and Hoffswell, Jane and Thazin Thane, Sao Myat and Bursztyn, Victor S. and Bearfield, Cindy Xiong},
	year         = 2025,
	journal      = {IEEE Trans. Vis. Comput. Graphics},
	volume       = 31,
	number       = 1,
	pages        = {536--546},
	doi          = {10/n82w},
    doiOiriginal = {10.1109/TVCG.2024.3456378}
}

@inproceedings{kafle2018:dvqa,
	title        = {{DVQA}: Understanding Data Visualizations via Question Answering},
	author       = {Kafle, Kushal and Price, Brian and Cohen, Scott and Kanan, Christopher},
	year         = 2018,
	month        = {June},
	booktitle    = {IEEE CVPR}
}

@inproceedings{savva2011:revision,
	title        = {{ReVision}: automated classification, analysis and redesign of chart images},
	author       = {Savva, Manolis and Kong, Nicholas and Chhajta, Arti and Fei-Fei, Li and Agrawala, Maneesh and Heer, Jeffrey},
	year         = 2011,
	booktitle    = {ACM Symposium on User Interface Software and Technology},
	publisher    = {ACM},
	series       = {UIST '11},
	pages        = {393--402},
	doi          = {10/bcbnfw},
    doiOiriginal = {10.1145/2047196.2047247}
}

@article{wu2024:rational,
	title        = {The Rational Agent Benchmark for Data Visualization},
	author       = {Wu, Yifan and Guo, Ziyang and Mamakos, Michalis and Hartline, Jason and Hullman, Jessica},
	year         = 2024,
	journal      = {IEEE Trans. Vis. Comput. Graphics},
	volume       = 30,
	number       = 1,
	pages        = {338--347},
	doi          = {10/hbjqpb},
    doiOiriginal = {10.1109/TVCG.2023.3326513}
}

@article{gelman2024:quartets,
	title        = {Causal Quartets: Different Ways to Attain the Same Average Treatment Effect},
	author       = {Andrew Gelman and Jessica Hullman and Lauren Kennedy},
	year         = 2024,
	journal      = {Am. Stat.},
	publisher    = {Taylor \& Francis},
	volume       = 78,
	number       = 3,
	pages        = {267--272},
	doi          = {10/qw5z},
    doiOiriginal = {10.1080/00031305.2023.2267597}
}

@article{fanelli2009:fabrication,
	title        = {How Many Scientists Fabricate and Falsify Research? A Systematic Review and Meta-Analysis of Survey Data},
	author       = {Fanelli, Daniele},
	year         = 2009,
	month        = {05},
	journal      = {PLOS ONE},
	publisher    = {Public Library of Science},
	volume       = 4,
	number       = 5,
	pages        = {1--11},
	doi          = {10/bn5pnj},
    doiOiriginal = {10.1371/journal.pone.0005738}
}

@article{alMarzouki2005:real,
	title        = {Are these data real? Statistical methods for the detection of data fabrication in clinical trials},
	author       = {Al-Marzouki, Sanaa and Evans, Stephen and Marshall, Tom and Roberts, Ian},
	year         = 2005,
	journal      = {BMJ},
	publisher    = {BMJ Publishing Group Ltd},
	volume       = 331,
	number       = 7511,
	pages        = {267--270},
	doi          = {10/btnnrv},
    doiOiriginal = {10.1136/bmj.331.7511.267},
}

@article{synder2024:divi,
	title        = {{DIVI}: Dynamically Interactive Visualization},
	author       = {Snyder, Luke S. and Heer, Jeffrey},
	year         = 2024,
	journal      = {IEEE Trans. Vis. Comput. Graphics},
	volume       = 30,
	number       = 1,
	pages        = {403--413},
	doi          = {10/qw54},
    doiOiriginal = {10.1109/TVCG.2023.3327172}
}

@article{liu2021:boba,
	title        = {Boba: Authoring and Visualizing Multiverse Analyses},
	author       = {Liu, Yang and Kale, Alex and Althoff, Tim and Heer, Jeffrey},
	year         = 2021,
	journal      = {IEEE Trans. Vis. Comput. Graphics},
	volume       = 27,
	number       = 2,
	pages        = {1753--1763},
	doi          = {10/ghgt7j},
    doiOiriginal = {10.1109/TVCG.2020.3028985}
}

@inproceedings{sarma2023:multiverse,
	title        = {{multiverse}: Multiplexing Alternative Data Analyses in R Notebooks},
	author       = {Sarma, Abhraneel and Kale, Alex and Moon, Michael Jongho and Taback, Nathan and Chevalier, Fanny and Hullman, Jessica and Kay, Matthew},
	year         = 2023,
	booktitle    = {ACM Human Factors in Computing Systems},
	publisher    = {ACM},
	series       = {CHI '23},
	doi          = {10/qw55},
    doiOiriginal = {10.1145/3544548.3580726},
	pages        = {148:1--15}
}

@inproceedings{saram2024:milliways,
	title        = {{Milliways}: Taming Multiverses through Principled Evaluation of Data Analysis Paths},
	author       = {Sarma, Abhraneel and Hwang, Kyle and Hullman, Jessica and Kay, Matthew},
	year         = 2024,
	booktitle    = {ACM Human Factors in Computing Systems},
	publisher    = {ACM},
	series       = {CHI '24},
	doi          = {10/qw56},
    doiOiriginal = {10.1145/3613904.3642375},
	pages        = {607:1--15}
}

@article{guo2025:vmc,
	title        = {{VMC}: A Grammar for Visualizing Statistical Model Checks},
	author       = {Guo, Ziyang and Kale, Alex and Kay, Matthew and Hullman, Jessica},
	year         = 2025,
	journal      = {IEEE Trans. Vis. Comput. Graphics},
	volume       = 31,
	number       = 1,
	pages        = {798--808},
	doi          = {10/qw57},
    doiOiriginal = {10.1109/TVCG.2024.3456402}
}

@misc{seeingTheory,
	title        = {Seeing Theory},
	author       = {Kunin, Daniel and Guo, Jingru and Dae Devlin, Tyler and Xiang, Daniel},
	year         = 2016,
	note         = {Last accessed Mar. 18, 2026. \url{https://seeing-theory.brown.edu/probability-distributions/index.html}}
}

@inproceedings{linsnic2024:fact,
	title        = {"Yeah, this graph doesn't show that": Analysis of Online Engagement with Misleading Data Visualizations},
	author       = {Lisnic, Maxim and Lex, Alexander and Kogan, Marina},
	year         = 2024,
	booktitle    = {ACM Human Factors in Computing Systems},
	publisher    = {ACM},
	series       = {CHI '24},
	doi          = {10/qhrc},
    doiOriginal  = {10.1145/3613904.3642448},
	pages        = {199:1--14}
}

@article{zhu2024:risk,
	title        = {A knowledge-guided visualization framework of disaster scenes for helping the public cognize risk information},
	author       = {Jun Zhu and Jinbin Zhang and Qing Zhu and Weilian Li and Jianlin Wu and Yukun Guo},
	year         = 2024,
	journal      = {International Journal of Geographical Information Science},
	publisher    = {Taylor \& Francis},
	volume       = 38,
	number       = 4,
	pages        = {626--653},
	doi          = {10/g4jdj9},
    doiOriginal  = {10.1080/13658816.2023.2298299}
}

@inproceedings{pan2024:markllm,
	title        = {{M}ark{LLM}: An Open-Source Toolkit for {LLM} Watermarking},
	author       = {Pan, Leyi  and Liu, Aiwei  and He, Zhiwei  and Gao, Zitian  and Zhao, Xuandong  and Lu, Yijian  and Zhou, Binglin  and Liu, Shuliang  and Hu, Xuming  and Wen, Lijie  and King, Irwin  and Yu, Philip S.},
	year         = 2024,
	booktitle    = {Proceedings of the 2024 Conference on Empirical Methods in Natural Language Processing: System Demonstrations},
	publisher    = {ACL},
	pages        = {61--71},
    doi          = {10/qw6f},
	doiOriginal  = {10.18653/v1/2024.emnlp-demo.7}
}

@article{brand2023:marketing,
	title        = {Using {LLM}s for Market Research},
	author       = {Brand, James and Israeli, Ayelet and Ngwe, Donald},
	year         = 2023,
	journal      = {Harvard Business School Marketing Unit Working Paper No. 23-062},
	pages        = {61--71},
    doi          = {10/qw6j},
	doiOriginal  = {10.2139/ssrn.4395751}
}

@article{braun2006:thematic,
	title        = {Using Thematic Analysis in Psychology},
	author       = {Braun, Virginia and Clarke, Victoria},
	year         = 2006,
	journal      = {Qualitative Research in Psychology},
	publisher    = {Routledge},
	volume       = 3,
	number       = 2,
	pages        = {77--101},
	doi          = {10/fswdcx},
    doiOriginal  = {10.1191/1478088706qp063oa}
}

@article{lake2017,
	title        = {Building machines that learn and think like people},
	author       = {Lake, Brenden M. and Ullman, Tomer D. and Tenenbaum, Joshua B. and Gershman, Samuel J.},
	year         = 2017,
	journal      = {Behav. Brain Sci.},
	volume       = 40,
	pages        = {e253:1--72},
    doi          = {10/gcw3n8},
	doiOriginal  = {10.1017/S0140525X16001837}
}

@article{livesey2017,
	title        = {Will human-like machines make human-like mistakes?},
	author       = {Livesey, Evan J. and Goldwater, Micah B. and Colagiuri, Ben},
	year         = 2017,
	journal      = {Behav. Brain Sci.},
	volume       = 40,
	pages        = {e270:41},
    doi          = {10/g537kf},
	doiOriginal  = {10.1017/S0140525X1700019X}
}
